\documentclass[%
aps,
rsi,%
 amsmath,amssymb,
 reprint,%
floatfix,
]{revtex4-1}
\usepackage{hyperref}
\usepackage{graphicx}% Include figure files
\usepackage{dcolumn}% Align table columns on decimal point
\usepackage{bm}% bold math
\usepackage[latin1,utf8]{inputenc}
\usepackage{float}
\usepackage{xcolor}
\usepackage{subfigure}
\usepackage{comment}
\usepackage[toc]{appendix}
\hypersetup{
     colorlinks = true,
     linkcolor = blue,
     anchorcolor = blue,
     citecolor = blue,
     filecolor = blue,
     urlcolor = blue
     }
\begin{document}

\preprint{AIP/123-QED}

\title{Verification of finite bath fluctuation theorem for a non-ergodic system}% Force line breaks with \\
%\thanks{Footnote to title of article.}

\author{Arthur M. Faria$^{1,}$}
\email{fariaart@ifi.unicamp.br}
\author{Marcus V. S. Bonança$^1$}
\affiliation{$^1$Instituto de Física Gleb Wataghin, Universidade Estadual de Campinas,   13083-859 Campinas, São Paulo, Brazil %\\This line break forced with \textbackslash\textbackslash
}%


\begin{abstract}

The analysis of fluctuations generated by a thermal reservoir has produced many results throughout the history of science, ranging from the verification of the atomic hypothesis, running through critical phenomena to the most recent advances in the description of non-equilibrium thermodynamic processes. Motivated by recent theoretical and experimental works, we analyze the non-equilibrium and equilibrium fluctuations caused by a finite and chaotic heat bath in a simple system of interest. Finite bath and system of interest give rise to a non-ergodic composite system when interacting with each other. We have characterized the equilibrium distribution induced by the finite bath and numerically verified the finite-bath fluctuation theorem. We have also verified the convergence of our results to Crooks' fluctuation theorem as the number of degrees of freedom of the finite bath increases while the non-ergodic character remains.

\end{abstract}

\maketitle

\section{\label{sec:level1}Introduction}

The fluctuations induced by a thermal reservoir, i.e., a system whose thermal capacity is infinitely large, are one of the most studied physical phenomena up to date. Its historical importance can be recognized in the formulation of the kinetic theory of gases \cite{maxwell, boltzmann}, in the verification of the atomic hypothesis through the study of the Brownian motion \cite{einstein} and, more recently, in the understanding of phase transitions \cite{stanley} and the non-equilibrium phenomena \cite{callen-welton, kubo1, kubo2, kubo-book, gallavotti-cohen, jarzynski1, jarzynski2, crooks1, crooks2, spohn, evans, Seifert_2012}. In all these cases, the thermal reservoir is considered to be a system that has, for all intents and purposes, an infinitely higher number of degrees of freedom than the system of interest itself.

Statistical mechanics has been successfully applied to described equilibrium fluctuations of macroscopic systems. However, recent theoretical and experimental developments have motivated the investigation of the fluctuations generated by finite baths \cite{smith, esposito, gorman, kantz, palzer, Esposito_2010, esposito1, bruderer, brantut, fialko, reeb, pekola, richens} extrapolating the standard description to the situation in which systems of few degrees of freedom are considered.

In this case, the thermal bath to which the system of interest is coupled to does not have an infinite heat capacity, and it is influenced by the coupling. However, it is well-known that phenomena such as dissipation and thermalization, usually attributed to the coupling to thermal reservoirs, can still be observed in this new regime \cite{wilkinson, berry1, berry, jarzynski, cohen, cohen1, bonanca-aguiar, marchiori-aguiar, xavier}. Another feature already reported in the literature is that the canonical distribution is no longer the one describing the equilibrium fluctuations. Under certain conditions, it can be shown that the equilibrium distribution of a system coupled to a finite bath is given by a power law \cite{khinchin, campisi2017_1,campisi2017_2,campisi2017_3}. Nevertheless, in the light of the heat theorem, originally introduced in the nineteenth century by Clausius, it is still possible to define thermodynamic quantities such as entropy on mesoscopic scales \cite{campisi2010, campisi2017_1, campisi2017_2, campisi2017_3, gallavotti}.

In the context of non-equilibrium fluctuations, most of the investigations since the pioneering work by Boltzmann on the transport equation \cite{boltzmann} have considered the coupling to standard reservoirs of heat and particles whose intensive thermodynamic state variables such as temperature and chemical potential remain always constant. However, some results recently reported have relaxed such conditions (see Ref. \cite{schaller_2014}). Among them, an analog of Crooks' fluctuation theorem has been derived for thermal baths of mesoscopic scales whose finite heat capacity must be energy-independent \cite{campisi_finite}. Additionally, system of interest plus thermal bath must be an \emph{ergodic} system \cite{lebowitz}.

In this work, we characterize the equilibrium and non-equilibrium fluctuations caused by a finite thermal bath of deterministic and chaotic dynamics when it is weakly coupled to a test system, and the total system is considered to be isolated and non-ergodic. In both cases, we vary the number of degrees of freedom of the bath, maintaining the non-ergodic character of the composite system. Regarding the non-equilibrium fluctuations, we verify an agreement with the Crooks fluctuation theorem as the number of degrees of freedom increases. 

This work is organized in the following way: in Section \ref{sec:level2}, we introduce the model adopted by pointing out the main aspects of it. Particularly, we present some properties of the bath that are necessary to observe the phenomenon of relaxation in the system of interest. The relaxation itself is investigated in Section \ref{sec:3}. Subsequently, in Section \ref{sec:4},  we characterize the equilibrium state of the system after the relaxation with the chaotic and finite heat bath. In the Section \ref{sec:5}, the system is driven out of equilibrium by two processes in order to investigate the fluctuation theorem for finite baths. 

 %In the limit of a large heat bath, the equilibrium state of the system is described by the Boltzmann-Gibbs distribution.  

%%%%%%%%%%%%%%%%%%%%%%%%%%%%%%%%%%%%%%%%%%%%%%%%%%%%%%%%%%%%%%%%%%%%%%%%%%%%%%%%%%%%%

\section{\label{sec:level2}Model}

We consider an isolated system governed by the Hamiltonian

    \begin{align}
	        \mathcal{H}=\mathcal{H}_{S}+\mathcal{H}_{I}+\mathcal{H}_{B},
	        \label{eq:hamilt}
    \end{align}

\noindent where

   \begin{align}
            \label{eq:HO}
	        &\mathcal{H}_{S} = \frac{P^2}{2} + \frac{k Q^2}{2},   \\
	        \label{eq:int}
	        &\mathcal{H}_{I} = \sum_{i=1}^{N}\lambda_N \,x_i\,Q = \sum_{i=1}^{N}\frac{\lambda}{\sqrt{N}}\, x_i\,Q , \\
	        &\mathcal{H}_{B} = \sum_{i=1}^{N}\frac{p_{x_i}^2 + p_{y_i}^2}{2} + \frac{x_i^2 y_i^2}{2} + a\frac{(x_i^4 + y_i^4)}{4}.
	        \label{eq:QS}
    \end{align}
    
The system of interest, whose dynamics is given by $\mathcal{H}_{S}$ in Eq. \eqref{eq:HO}, is a one-dimensional harmonic oscillator (HO) with phase-space variables $Q$, $P$, and spring constant $k$. The thermal bath is modeled as a collection of non-interacting and identical two-dimensional quartic systems (QS) (see Eq. \eqref{eq:QS}), whose microscopic states are given by the phase-space variables $(x_{i},y_{i},p_{x_{i}},p_{y_{i}})$, with $i=1,\cdots,N$. The parameter $a$ is the same for all of them and assumes values between $0$ and $1$. The interaction between the HO and the thermal bath is given by the Hamiltonian $\mathcal{H}_{I}$, Eq. \eqref{eq:int}, which is essentially a bilinear coupling between the position coordinates $x_i$ and $Q$. The coupling constant $\lambda_N$ depends on the number $N$ of QS in the thermal bath. We use $\lambda_N = \lambda/\sqrt{N}$ so that the effective coupling between the HO and the thermal bath becomes independent of $N$. This is so because the back-action on the HO due to the coupling is in leading order quadratic in $\lambda_N$ (see Ref. \cite{marchiori-aguiar} for more details).

We can intuitively conclude that the dynamics of the complete system will be \emph{non}-ergodic only by looking at the Hamiltonians \eqref{eq:hamilt} to \eqref{eq:QS}. Large portions of the whole phase-space will be hardly explored due to the absence of direct interaction among the several QS and the weak interaction between the HO and each QS. Furthermore, it was already shown that the Hamiltonian of a single QS always has stable orbits \cite{dahlqvist1990}. However, the dynamics of QS can be considered effectively ergodic (in the sense of Ref. \cite{khinchin}) for certain values of the parameter $a$.

\subsection{Properties of the QS Hamiltonian}
\begin{figure}[!ht]
	\centering
    \includegraphics[width=0.48\textwidth]{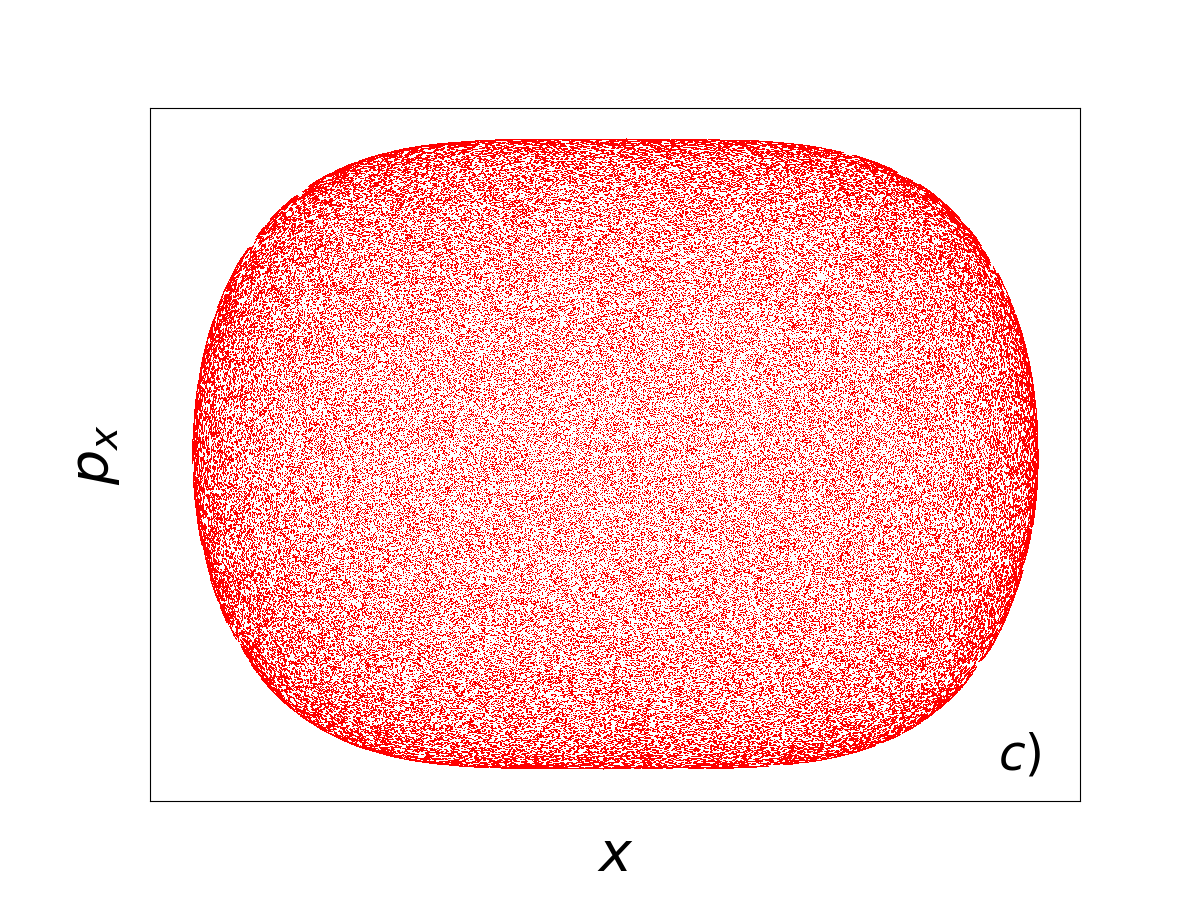}
    \caption{Poincaré sections of a QS when the plane $ y = 0 $ and the condition $ p_y> 0$ are considered for different values of the $ a $ parameter: (a) $a = 1$, (b) $a = 0.5$, (c) $a = 0.1$. These sections were obtained using the symplectic numerical integrator of Ref.~\cite{forest}, with 100 initial conditions sampled microcanonically with energy $E_{QS}=0.5$.}
      \label{fig:poinc}
\end{figure}

We will now comment on the properties of a single QS. Each QS Hamiltonian (defined by the terms under the summation sign in Eq.~(\ref{eq:QS})) is invariant under a scaling transformation of momenta and position coordinates so that the dynamics in different energy shells are qualitatively the same and only change by a scaling factor \cite{percival}. This property allows us to vary the QS’s energy and time scales without changing its dynamical regime. In fact, the dynamics is only controlled by the parameter $a$ in such way that the QS is integrable for $a = 1.0$, effectively chaotic for $a= 0.1$ and mixed for intermediate values. This behavior can be observed using Poincaré's surfaces of sections. The Poincaré's section is a two-dimensional plane that sections the surface of constant energy so that the trajectories of the system intersecting it construct a map. For systems of two degrees of freedom, the dynamic properties of the original four-dimensional phase space are accurately reflected in this map \cite{lichtenberg}. We see that, for $a = 1$ (see Fig.\ref{fig:poinc}\color{blue}a\color{black}), all trajectories present a complete regular behavior since energy and angular momentum are conserved. In this case, one can observe a clear distinction between trajectories with positive and negative values of angular momentum. On the other hand, for $a = 0.5$, Fig. \ref{fig:poinc}\color{blue}b\color{black})  shows a mixture of regular and irregular behavior depicted in the panel by smooth and dotted structures. Finally, for $a = 0.1$ (see Fig. \ref{fig:poinc}\color{blue}c\color{black}), the trajectories display approximately ergodic behavior since the phase space points seem to be uniformly distributed over the section. 

\begin{figure}[!ht]
	\centering
	\includegraphics[width=0.5\textwidth]{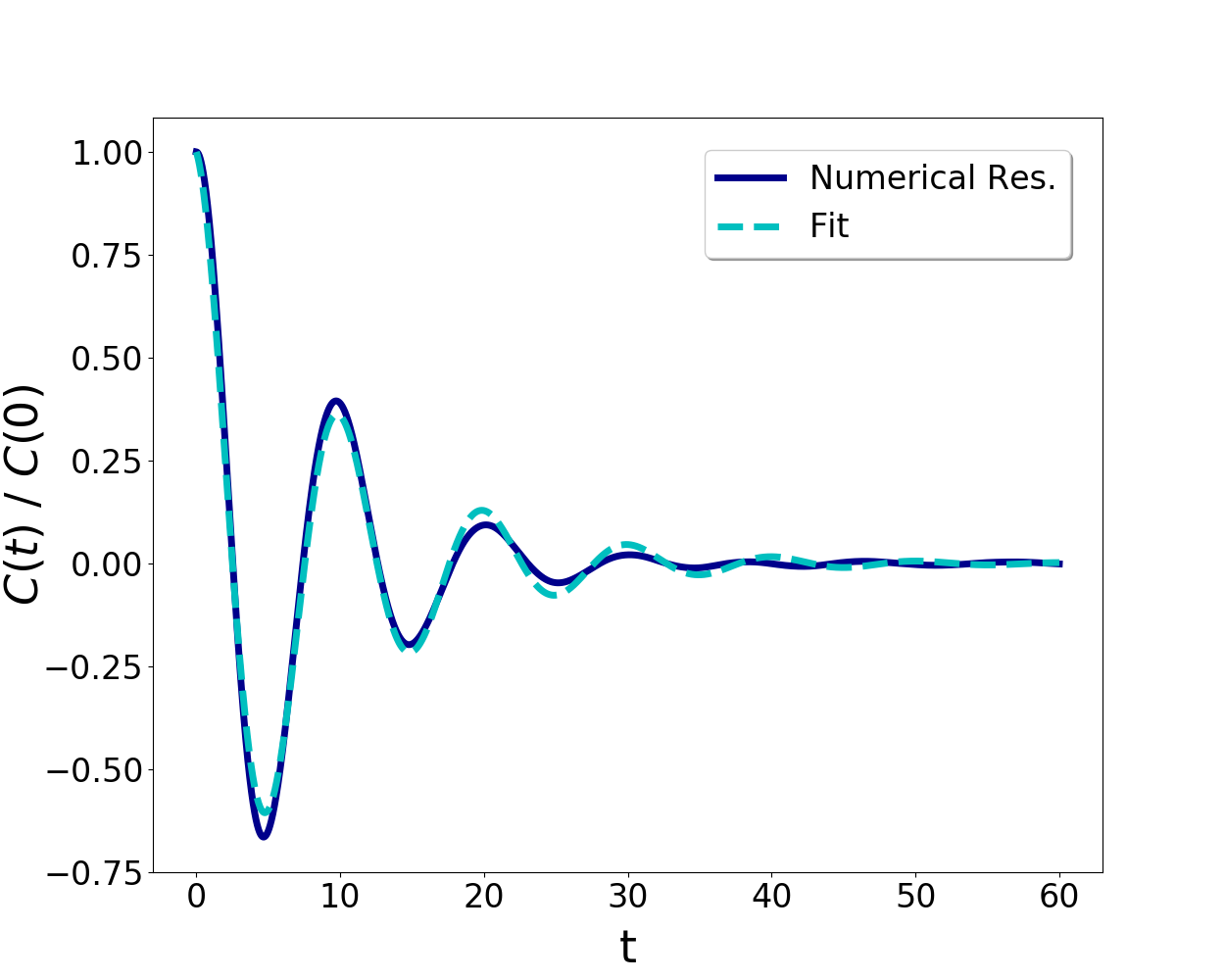}
	\caption{Normalized correlation function, $ \langle x_{i}(0) x_{i}(t) \rangle / \langle x_{i}^2 (0) \rangle $, for the energy $ E_{QS} = 0.5 $. The solid line corresponds to the numerical result and the dashed line to the fit $C(t) = C(0)e^{-\alpha \,t}\cos{(\gamma \,t)}$, where $\alpha$ and $\gamma$ are free parameters and $C(0)=\langle x_{i}^{2}\rangle$. We used $a = 0.1$ and $ 10 ^ 6 $ initial conditions sampled microcanonically.}
	\label{fig:corr}
\end{figure}

The chaoticity of QS for $a=0.1$ can be also verified through the behavior of correlation functions. We have calculated them numerically using symplectic algorithms to integrate Hamilton's equations \cite{forest}. We have done that sampling several initial conditions with the same energy according to a microcanonical distribution. The result for the autocorrelation function $ \langle x_{i}(0) x_{i}(t) \rangle $ is shown in Fig. \ref{fig:corr} together with a fitting by an exponential decay.  After a certain time, it is expected that the correlation function tends to an equilibrium average value. In this case, in particular, the average of $x_{i}$ over a microcanonical distribution is zero since the QS Hamiltonian remains invariant when $x_{i}\to -x_{i}$. In other words,

   \begin{align}
        \lim_{t \to \infty}\langle x_{i}(0)x_{i}(t)\rangle \to \langle x_{i} \rangle^2 = 0,
    \end{align}

\noindent where $\langle\cdot\rangle$ denotes the microcanonical average.

%%%%%%%%%%%%%%%%%%%%%%%%%%%%%%%%%%%%%%%%%%%%%%%%%%%%%%%%%%%%%%%%%%%%%%%%%%%%%%%%%%%%%%%%%%%%%%%%%%%%%%

%%%%%%%%%%%%%%%%%%%%%%%%%%%%%%%%%%%%%%%%%%%%%%%%%%%%%%%%%%%%%%%%%%%%%%%%%%%%%%%%%%%%%%%%%%%%%%%%%%%%%%

\begin{figure}[!ht]
	\centering
	\includegraphics[width=0.5\textwidth]{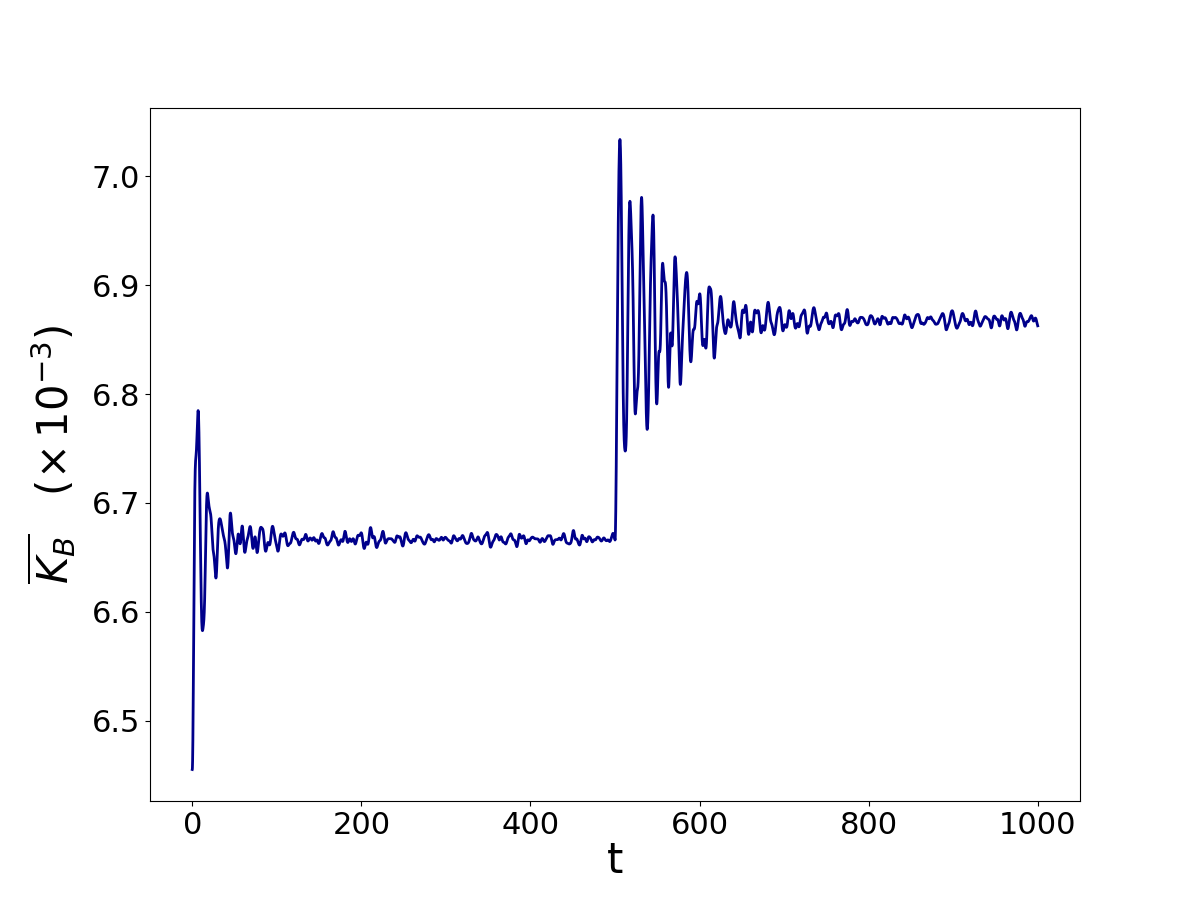}
	\caption{Time evolution of the ensemble average of $ K_{QS} = (p_{x_{i}}^2 + p_{y_{i}}^2)/2 $ when the the value of $a$ jumps from $ 0.1 $ to $ 0.12$. We used $ 10^6 $ initial conditions, an initial energy $ E_{QS} = 0.01 $, and $ a = 0.1 $ initially. After 500 units of time, we changed the parameter $a$ abruptly to $ 0.12 $. We let then the average of $K_{QS}$ relax for another 500 units of time.}
	\label{fig:relax}
\end{figure}

Another indication of the QS chaoticity when $a\ll 1.0$ is the self-relaxation shown in Fig.\ref{fig:relax}. In this case, we calculated numerically the time evolution of the average value of the kinetic energy, $ K_{QS} = (p_{x_{i}}^2 + p_{y_{i}}^2)/2$, first by evolving initial conditions sampled from an approximately microcanonical distribution with $a = 0.1$ and energy value $E_{QS}=0.01$ fixed. We observe then that the average value relaxes to a stationary value corresponding the microcanonical average (see Appendix \ref{app:A}). After a certain time interval, the value of $a$ was abruptly changed to $a = 0.12$ and the average of $K_{QS}$ once more relaxes to a new microcanonical average. 

 \begin{figure}[!ht]
	\centering
    \includegraphics[width=0.48\textwidth]{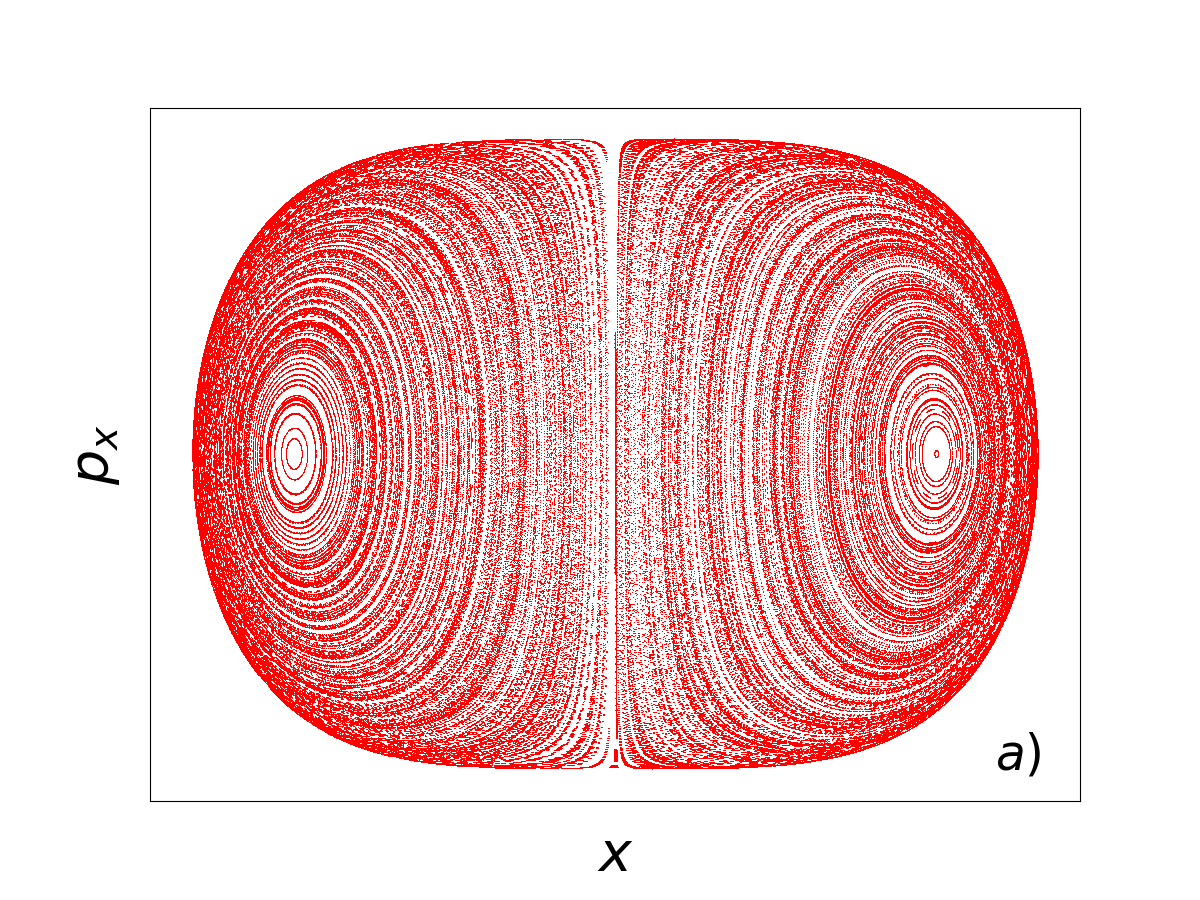}	
\end{figure}
\begin{figure}[!ht]
	\centering
    \includegraphics[width=0.48\textwidth]{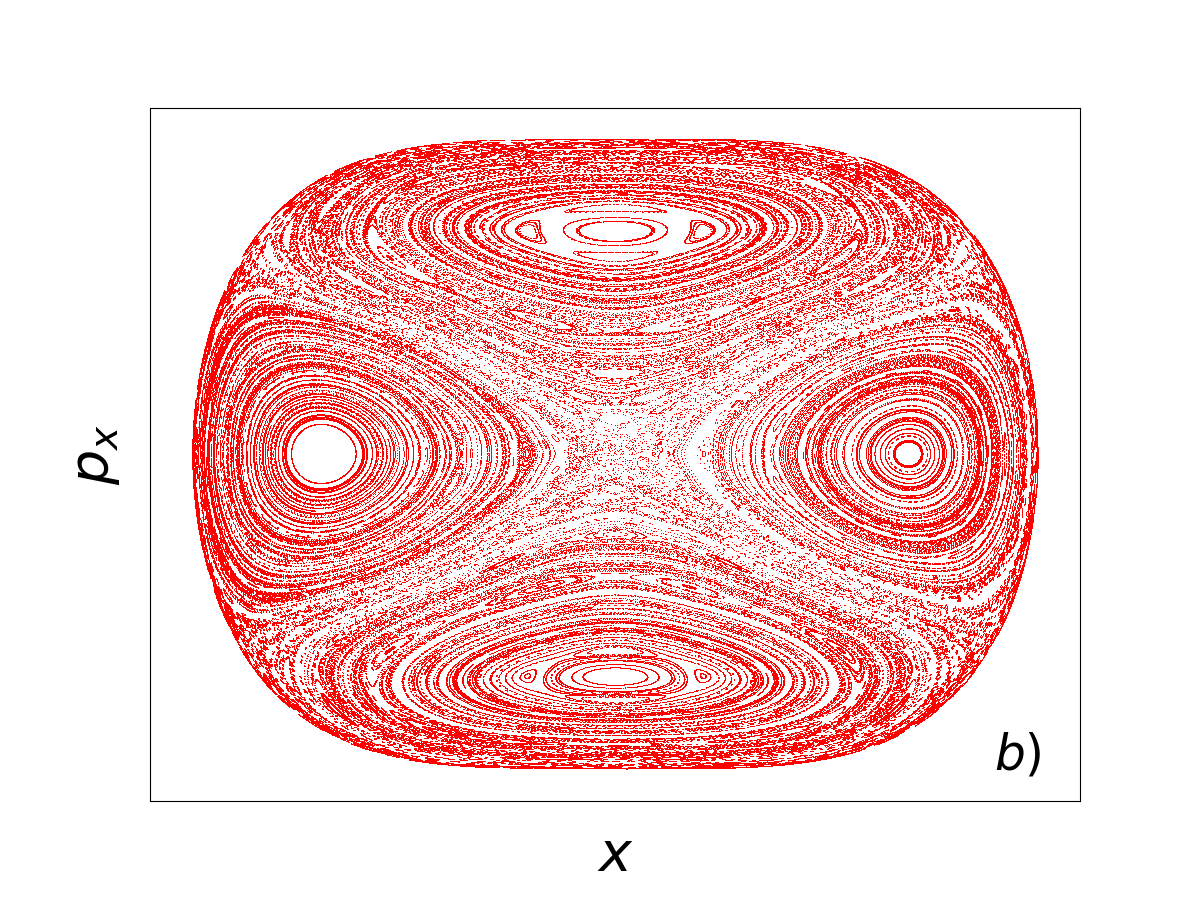}
\end{figure}

\section{\label{sec:3}Relaxation due to a finite heat bath}

To verify the fluctuation theorem for the model described previously, we need first to find the equilibrium distribution of HO due to the coupling to a collection of QS. However, it is not evident that such equilibrium state will exist. Hence, we will first show in this section that our model does reach equilibrium and, in the next section, we present numerical results that corroborate our analytical expression for the equilibrium distribution.

We go back then to the model \eqref{eq:hamilt} and investigate the relaxation between the HO and a collection of QS when $a=0.1$. For large values of $N$, we guarantee that each QS is weakly coupled to the HO. On the other hand, we have a large number of QS coupled to the HO. Thus, the coupling strength between the HO and the collection of QS is indeed controlled by the value of $\lambda$ in 
Eq.~\eqref{eq:int}.

As our starting point, we derive an analytic criterion to determine that a system has reached the equilibrium when weakly coupled to a heat bath composed of $N$ non-interacting sub-systems. Assuming that the interaction energy is negligible, the phase-space volume $\Omega_{tot}(E_{tot})$ of the \emph{total} system enclosed by a surface of constant energy $E_{tot}$ is given by \cite{khinchin}

\begin{align}
             \Omega_{tot}(E_{tot}) &= \int_{\mathcal{H}_{tot} \leq E_{tot}} dz_{tot}\,   \\
             &=   \int_{\mathcal{H}_{S} \leq E_{tot}} dz\,
             \int_{\mathcal{H}_{B} \leq E_{tot} - \mathcal{H}_S} dz_B   \\
             &=  \int_{\mathcal{H}_{S} \leq E_{tot}} \Omega_B( E_{tot} - \mathcal{H}_S) dz, 
        \end{align}
        
\noindent where $z_{tot}$, $z$ and $z_{B}$ denote respectively the canonical phase-space variables of the total system, of the system of interest and of the heat bath. The last expression can be written as \cite{khinchin}

\begin{align}
    \Omega_{tot}(E_{tot}) = \int_{0}^{E_{tot}} \omega_S(\mathcal{H}_S)\,\Omega_B(E_{tot} - \mathcal{H}_S)\, d\mathcal{H}_S,
    \label{eq:conv}
\end{align}

\noindent where the density of states (DOS) $\omega_{S}(E)$ is defined as $\omega_{S}(E) = \partial \,\Omega_{S}(E)/\partial E$, $\mathcal{H}_{S}$ is the Hamiltonian of the system of interest and $\Omega_{B}(E_{B})$ is the phase-space volume enclosed by the surface $\mathcal{H}_{B}(z_{B})=E_{B}$ where $\mathcal{H}_{B}$ is the Hamiltonian of the heat bath.

As mentioned before, the $N$ sub-systems in the heat bath do not interact with each other and $\Omega_B(E_{tot} - \mathcal{H}_S)$ can be further expressed as \cite{khinchin} 

\begin{align}
    &\Omega_B(E_{tot}- \mathcal{H}_S) =\nonumber\\ &\int_{0}^{E_B}d\mathcal{H}_{B_{1}} \int_{0}^{E_B-\mathcal{H}_{B_{1}}}d\mathcal{H}_{B_{2}}\ldots\int_{0}^{E_B-\sum_{i=1}^{N-2}\mathcal{H}_{B_{i}}}d\mathcal{H}_{B_{N-1}} \nonumber \\ &\left(\prod_{i=1}^{N-1}\omega_{B_i}(\mathcal{H}_{B_i}) \right) \Omega_{B_{N}}\left(E_{tot} - \mathcal{H}_S - \sum_{i=1}^{N-1}\mathcal{H}_{B_{i}}\right),
    \label{eq:bath_conv}
\end{align}
where we denote by $\mathcal{H}_{B_{i}}$ the Hamiltonian of the $i$th sub-system in the heat bath. Once more, the assumption of weak coupling implies that $E_{tot} = \mathcal{H}_S + \mathcal{H}_B$ so that $\mathcal{H}_B = \sum_{i=1}^{N}\mathcal{H}_{B_i}  = E_{tot} - \mathcal{H}_S$.

In fact, when both system and heat bath are macroscopic, the equilibrium between them is reached once their temperatures are equal, i.e.,
\begin{align}
    T_S = T_B,
    \label{cu123}
\end{align}

\noindent with each temperature related to its corresponding entropy $\mathcal{S}$ through the expression, $T^{-1} = \partial \mathcal{S}/ \partial E$. We will assume that condition (\ref{cu123}), which is a consequence of the maximal-entropy principle, also holds when the systems are not in the thermodynamic limit \cite {hanggi} and we will obtain their temperatures using the Gibbs entropy \cite {hanggi, campisi2015}, namely, $\mathcal{S}=  k_\mathcal{B}\ln{\Omega(E)}$, where $ k_\mathcal{B}$ is the Boltzmann constant and $\Omega(E)$ is the phase-space volume enclosed by a given energy shell. 

We consider from now on that both phase-space volumes, $\Omega_{S}$ and $\Omega_{B_{i}}$ are power laws of the energies (see Appendix \ref{app:B}),

\begin{align}
    \Omega_S(E_S) \propto E_S^{\, \kappa_{S}},\qquad\Omega_{B_i}(E_{B_i}) \propto E_{B_i}^{\, \kappa_{B_{i}}}.
    \label{eq:vol_phase-spa}
\end{align}

\noindent The Gibbs entropy, $\mathcal{S} = k_{\mathcal{B}}\ln \Omega$, then gives us that

\begin{align}
    \frac{1}{T_{\alpha}} = \frac{\partial \mathcal{S}_{\alpha}}{\partial E_{\alpha}} = k_{\mathcal{B}}\frac{\partial \ln\Omega_{\alpha}}{\partial E_{\alpha}} = k_{\mathcal{B}}\frac{\omega_\alpha}{\Omega_\alpha} = k_{\mathcal{B}} \frac{\kappa_{\alpha}}{E_{\alpha}},
\end{align}

\noindent where $\alpha$ can be either $S$ or $B_{i}$. The quantities $\kappa_S$ and $\kappa_{B_{i}}$ are then clearly related to the microcanonical heat capacities defined as follows (see Appendix \ref{app:B})

\begin{align}
        C_\alpha = \left(\frac{\partial}{\partial E_\alpha}T_\alpha(E_\alpha)\right)^{-1} = k_{\mathcal{B}}\, \kappa_{\alpha}.
        \label{cuenver}
    \end{align}

\noindent Furthermore, we also notice that
    \begin{align}
        T_\alpha(E_\alpha) &= \frac{E_\alpha}{C_\alpha}.
        \label{eq:temp_cap}
    \end{align}

\noindent The phase-space volumes $\Omega_{S}$ and $\Omega_{B}$ can be obtained analytically when the HO and the QS are considered as the system of interest and a heat bath component, respectively (see Appendix \ref{app:B} for further information about $\Omega_{S}$ and $\Omega_{B}$). In particular, we obtain $C_S/k_{\mathcal{B}} = 1 $ for the HO and $C_{B}^{(1)}/k_{\mathcal{B}} = 3/2 $ for a single QS.

Condition (\ref{cu123}) together with Eqs.~ (\ref{cuenver}) and (\ref{eq:temp_cap}) lead then to the following relation between the equilibrium energies $E_S^{(eq)}$ of the system and $E_B^{(eq)}$ of heat bath for the model described by (\ref{eq:hamilt}) to (\ref{eq:QS}),

\begin{align}
   \mathcal{R} = \frac{E_S^{(eq)}}{E_B^{(eq)}} = \frac{C_{S}}{C_{B}} = \frac{2}{3N}.
   \label{eq:cond}
\end{align}

\begin{figure}[!ht]
	\centering
	\includegraphics[width=0.5\textwidth]{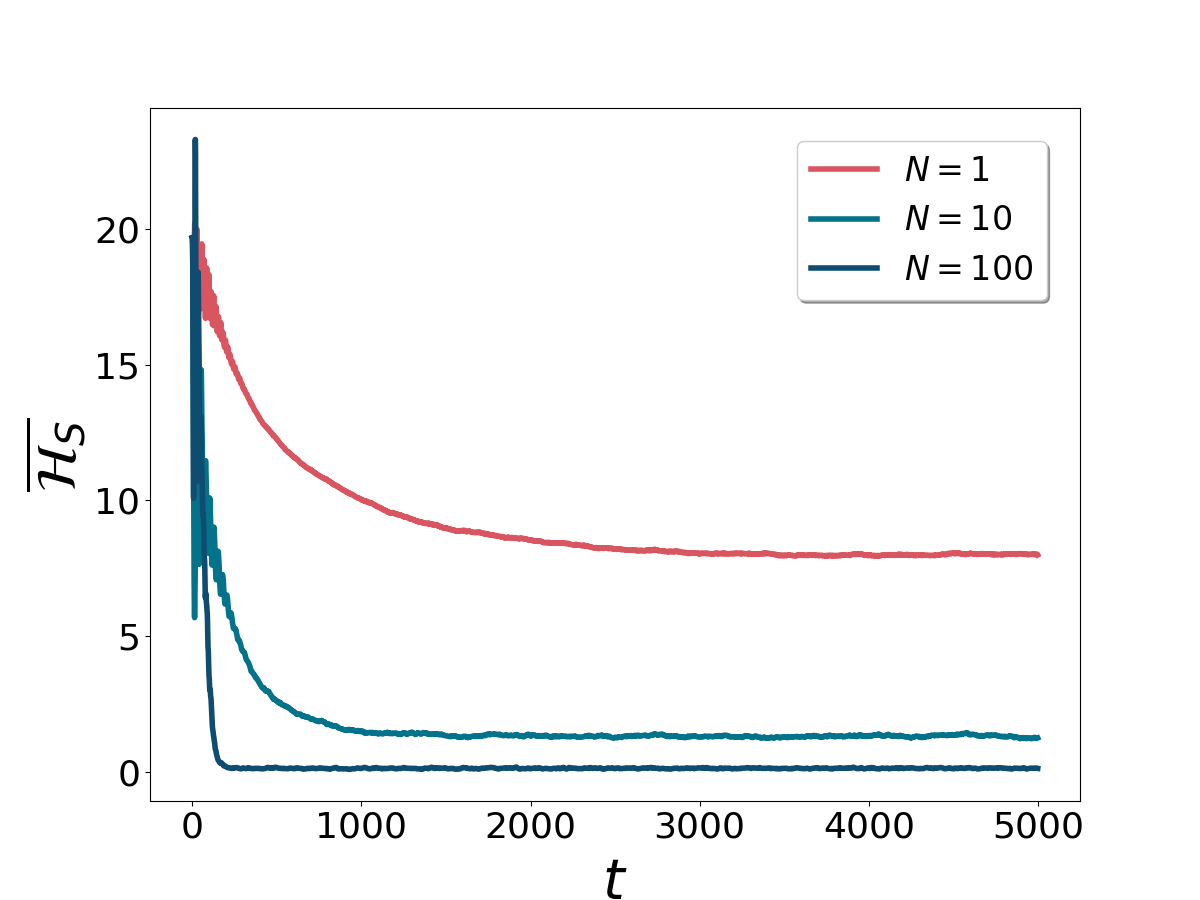}% Here is how to import EPS art
	\caption{Relaxation of the HO average energy $\overline{\mathcal{H}}_S$ (see Eq.~(\ref{eq:HO})) for different numbers of QS in the heat bath. The system and bath initial conditions were sampled microcanonically with the respective energies $E_S=19.7$, and $E_B=10^{-5}$. The red $(N=1)$, green ($N=10$) and blue ($N=100$) curves were obtained with $10^5$, $10^2$ and $10$ initial conditions, respectively. As the number of degrees of freedom increases, wide fluctuations are smoothed out and a smaller number of initial conditions is required to obtained good average values. The numerical results were obtained with $a=0.1$, $k=0.3$ and $\lambda=0.1$.}
	\label{fig:dissip}
\end{figure}

\noindent As in Appendix \ref{app:B}, we denote here by $ C_{B} = N C_{B}^{(1)} $ the total heat capacity of the $N$ non-interacting QS. Figure~\ref{fig:dissip} shows the relaxation of HO energy for three different values of $N$. Numerical and analytical results for the corresponding ratios between the equilibrium energies are shown in the Table \ref{tab:table1}. 

\begin{table}[!ht]
	\begin{ruledtabular}
		\begin{tabular}{cccc}
			 $ \mathcal{R}$             & $N=1$  & $N=10$ & $N=100$ \\
			\hline
			\\
             \textbf{An.}  & 0.67    & 0.067    & 0.0067   \\
			 \textbf{Num.}  & $0.677 \pm 0.002$    & $(6.77  \pm 0.02)10^{-2} $   &   $(6.55  \pm 0.02)10^{-3} $     \\
		\end{tabular}
		\caption{Numerical (Num.) and analytical (An.) results of the HO and heat bath equilibrium energy ratio $\mathcal{R}$ (\ref{eq:cond}) for $N=1$, $N=10$, and $N=100$. The numerical results were extracted from Fig.~\eqref{fig:dissip}.}
		\label{tab:table1}
	\end{ruledtabular}
\end{table}
   
Condition \eqref{eq:cond} must be satisfied in the equilibrium state when the interaction between system and heat bath is extremely weak. The good agreement between numerical and analytical results shown in Table \ref{tab:table1} is an indication that the value chosen for $\lambda$ in fact leads to a weak coupling regime.

   %***************Hist-Energ***********************

\section{ \label{sec:4}Equilibrium distribution due to a finite bath}

The equilibrium distribution of a system of interest weakly coupled to a heat bath is given by the following probability distribution function (PDF) when the total system is ergodic \cite{khinchin,campisi_finite}

\begin{align}
    \rho(z,k) = \frac{\omega_B(E_{tot}-\mathcal{H}_S(z,k))}{\omega_{tot}(E_{tot})},
    \label{eq:pdf_ergodic}
\end{align}

\noindent where $z$ denotes the phase-space variables and $\omega_B(E_{tot}-\mathcal{H}_S(z,k))$ is the bath DOS evaluated at energy $E_{tot}-\mathcal{H}_S(z,k)$. When the heat bath is a collection of $N$ non-interacting components, the total system is certainly not ergodic (see Sec.~\ref{sec:level2}). However, we will show that, surprisingly, Eq.~(\ref{eq:pdf_ergodic}) remains valid for our model.

The DOS $\omega_B(E_B)$ of our finite bath is obtained deriving Eq.~\eqref{eq:bath_conv} with respect to $E_B$. If the DOS of each component is a power-law then Eq.~(\ref{eq:bath_conv}) leads to a $\omega_{B}(E_{B})$ which is also a power-law (see Appendix~\ref{app:B}). In this case, it is possible to rewrite the equilibrium PDF as \cite{campisi_finite,campisi2017_1}

\begin{align}
	   \rho(z, U_S, k) = \mathcal{N}^{-1}\left[1-\frac{(\mathcal{H}_S(z,k) - U_{S})}{C_B T}\right]_+^{\frac{C_B}{k_\mathcal{B}} -1},
	   \label{eq:eq_est}
\end{align}

\noindent where $C_{B}$ is the finite-bath heat capacity (see Eq.~(\ref{cuenver})) and the normalization factor $\mathcal{N}$ is the following function of $U_{S}$ and $k$,

\begin{align}
	   \mathcal{N}(U_S, k) = \int dz \left[1-\frac{(\mathcal{H}_S(z,k) - U_{S})}{C_BT}\right]_+^{\frac{C_B}{k_\mathcal{B}} -1}.
	   \label{eq:norm}
\end{align}

The quantity $U_S$ represents the internal energy of the system and $T$ is the equilibrium temperature between system and bath. They are defined as 

\begin{align}
    U_S \doteq \langle \mathcal{H}_S\rangle_{\rho},\;\;\mathrm{and}\;\;T \doteq \langle K_S\rangle_{\rho},
     \label{eq:US}
\end{align}

\noindent where $\langle\cdot\rangle_{\rho}$ denotes the phase-space average using the PDF (\ref{eq:eq_est}) and $K_S$ is the kinetic-energy part of $\mathcal{H}_S$. Equations~(\ref{eq:US}) can be understood as implicit equations either for $U_S(T,k)$ or $T(U_{S},k)$. The total energy in equilibrium is $ E_{tot} = U_S + C_B T $ where $C_B T$ represents the finite bath internal energy $U_B$ (see Eq.~\eqref{eq:temp_cap}). Additionally, the symbol $[x]_{+}$ is defined as $[x]_{+} = x \Theta(x)$ with $\Theta(x)$ denoting the Heaviside step function. This accounts for the sharp cutoff that must exist in the PDF (\ref{eq:eq_est}) when the system's energy reaches $E_{tot}$. It can be shown that Eq.~(\ref{eq:eq_est}) furnishes a statistical-mechanical description as valid as those obtained from the well-known equilibrium ensembles. We refer to \cite{campisi2017_1,campisi2017_2,campisi2017_3} for more details.

\begin{figure}[!ht]
    \centering
    \includegraphics[width=0.50\textwidth]{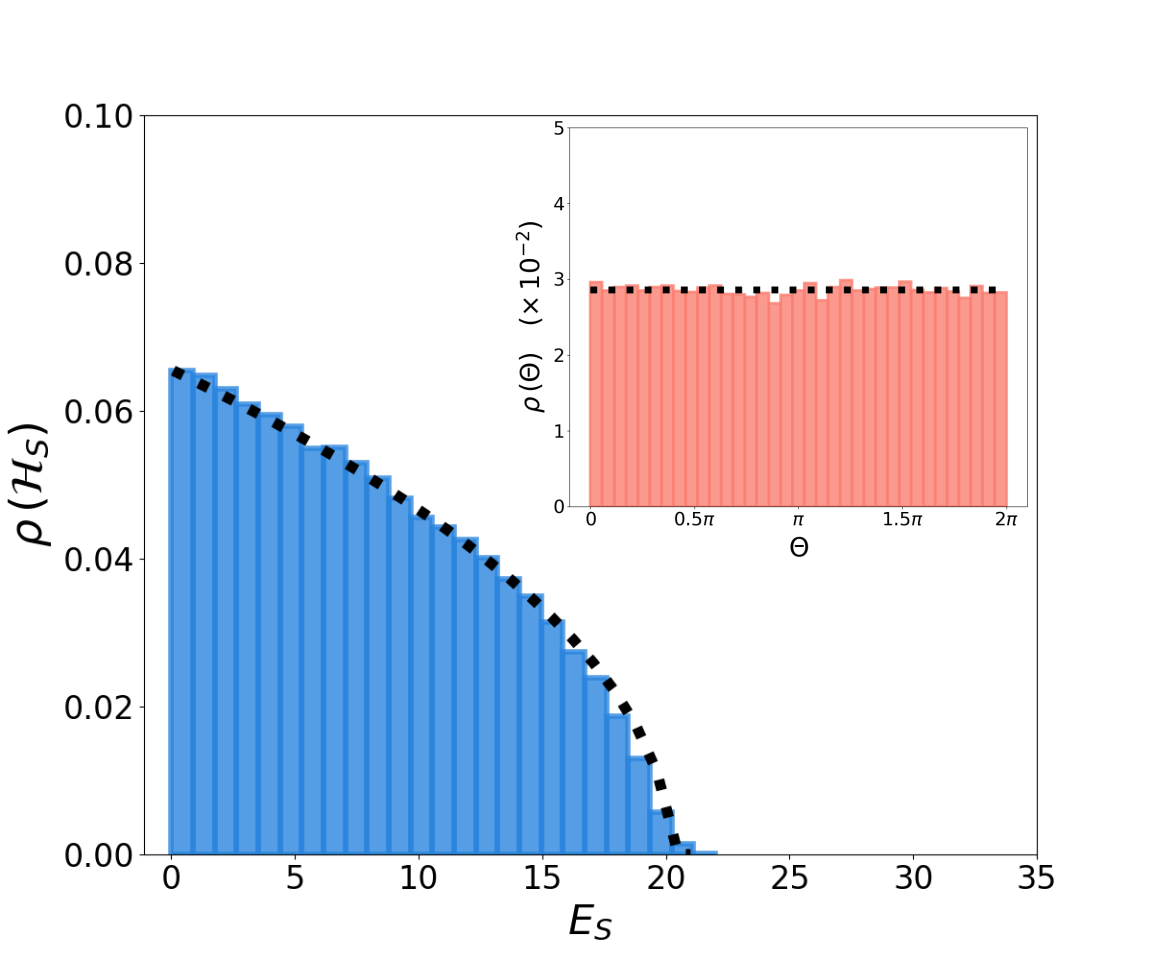}% Here is how to import EPS art
	\caption{Distributions of energy and angle of the HO in equilibrium with a single QS $(N=1)$. The figure shows the energy histogram (blue bar histogram), and the fit \eqref{yu1} (black dotted line). Inset: distribution of angles (red bar histogram), and the fit (black dotted line) $f(\theta) = b\theta+c$. We sampled microcanonically $10^5$ initial conditions with $E_S=19.7$, $E_B=10^{-5}$, $a=0.1$, $k = 0.3$, and $\lambda=0.1$.}
	\label{qu}
\end{figure}

\begin{figure}[!ht]
    \centering
	\includegraphics[width=0.5\textwidth]{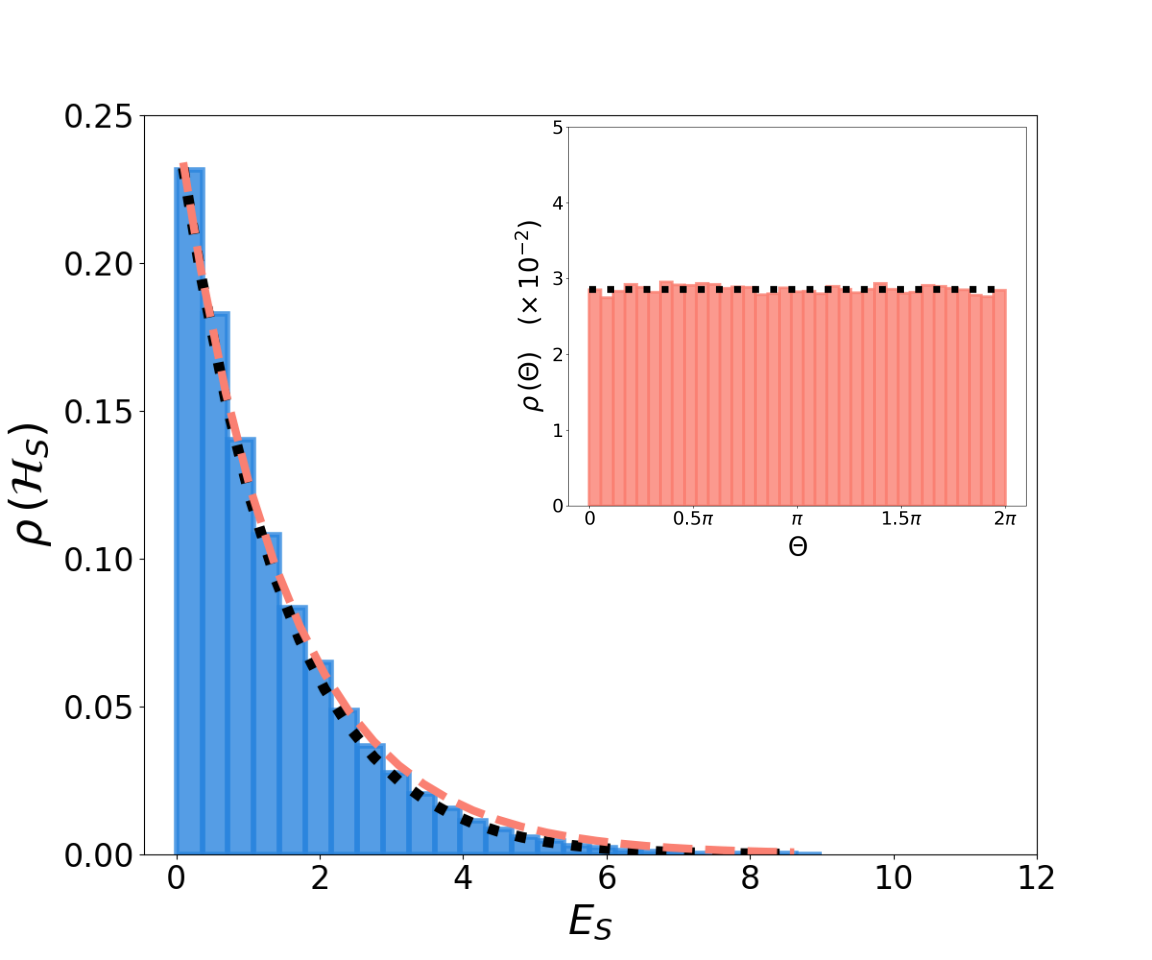}% Here is how to import EPS art
	\caption{Distributions of energy and angle of the HO in equilibrium with ten QS $(N=10)$. The figure shows the energy histogram (blue bar histogram), the fit \eqref{yu1} (black dotted line), and the fit of a canonical distribution (salmon dashed line). Inset: distribution of angles (red bar histogram), and the fit (black dotted line) $f(\theta) = b\theta+c$. We sampled microcanonically $10^5$ initial conditions with $E_S=19.7$, $E_B=10^{-5}$, $a=0.1$, $k = 0.3$, and $\lambda=0.1$.}
	\label{qu2}
\end{figure}

We have verified numerically that, for the model given by Eqs.~\eqref{eq:HO} to (\ref{eq:QS}), the equilibrium distribution of the HO indeed follows Eq.~(\ref{eq:eq_est}) (see Figs.~\ref{qu} and \ref{qu2}). As shown in Fig.~\ref{fig:dissip}, the average value $ \overline{\mathcal{H}}_S $ reaches a steady value after approximately $ 5000$ units of time (u.t.), suggesting that the HO is in its equilibrium state at that moment. Thus, we defined $\tau_S =5000$ u.t. as the relaxation time of the HO. To obtain the HO energy distribution numerically, we have simulated the contact between the HO and its heat bath (\ref{eq:QS}) up to the time interval $\tau_S $. The energies sampled from the ensemble of trajectories evolved up to $\tau_S$ are shown in the histograms of Figs.~\ref{qu} and \ref{qu2} for two different values of $N$. Both figures also show the fitting of the histograms by the expression 

    \begin{align}
	       f(\mathcal{H}_{S}, k) = A \left(1- \frac{1}{C_B T} \,(\mathcal{H}_{S} - U_S)\right)^{\frac{C_B}{k_{\mathcal{B}}}-1}, 
	       \label{yu1}
    \end{align}

\noindent where $A$ as the only free parameter and the values of $U_S$ and $T$ taken from the relaxation curves of Fig.~\ref{fig:dissip}. The value of $C_B/k_{\mathcal{B}}$ was taken equal to its analytical prediction, namely, $C_B/k_{\mathcal{B}}= N C_B^{(1)}/k_{\mathcal{B}}=3N/2$. One notices a good agreement between the analytical expression and the numerical histograms.

The insets of Figs.~\ref{qu} and \ref{qu2} show the angular distribution obtained numerically. The angle $\theta$ is defined according to the parameterization $P = \sqrt{2 \mathcal{H}_{S}}\sin\theta$, $Q = \sqrt{2 \mathcal{H}_{S}/k}\cos\theta$, where $P$ and $Q$ are the momentum and position appearing in the HO Hamiltonian (\ref{eq:HO}). The expression (\ref{yu1}) is proportional to the marginal distribution  $\rho\left(\mathcal{H}_S,k\right)$ of \eqref{eq:eq_est}. 
The other marginal distribution, $\rho (\theta, k)$, is obtained as follows  

    \begin{align}
        \rho(\theta,k) = \int_{0}^{\infty} \rho\left(\mathbf{z}(\mathcal{H}_S, \theta),k\right) J(\mathcal{H}_S, \theta) \, d\mathcal{H}_S,
    \end{align}

\noindent where $J(\mathcal{H}_{S}, \theta)$ denotes the Jacobian of the transformation $(P,Q)\to(\mathcal{H}_{S},\theta)$. Since Eq.~\eqref{eq:eq_est} and $J(\mathcal{H}_{S}, \theta)$ are independent of $\theta$ for the HO, the marginal distribution $ \rho(\theta,k)$ must be uniform. In the inset of Fig.~\ref{qu}, we have the numerical result for this marginal considering $N=1$. 
A fit given by the linear expression $f(\theta) = b\theta+c $ gives $b = 0$ and $c =0.03$ (similar results were obtained for the inset of Fig.~(\ref{qu2})). Thus, the two marginals confirm that the equilibrium statistics of the HO does follow Eq.~(\ref{eq:eq_est}) for our model.

It can be shown that the distribution \eqref{eq:eq_est} interpolates between the canonical, $\rho_{c}$, and the microcanonical, $\rho_{mc}$, PDFs as the heat capacity $C_{B}$ of the heat bath varies \cite{campisi_finite}. In other words, Eq.~\eqref{eq:eq_est} leads to
    \begin{align}
        \label{eq:canon}
        &\lim_{C_B/k_{\mathcal{B}} \to \infty}\rho = \rho_{c} = \frac{e^{-\beta\mathcal{H}_S(\mathbf{z},k)}}{\mathcal{Z}(\beta,k)}, \\
	   &\lim_{C_B/k_{\mathcal{B}} \to 0}\rho = \rho_{mc} = \frac{\delta(U_S - \mathcal{H}_S(\mathbf{z},k))}{\omega(U_S,k)}.
    \end{align}

\noindent In particular, for $N=10$, the DOS obtained via Eq.~\eqref{eq:bath_conv}, $\omega_B(E_B) \propto E_B^{14}$, gives $C_{B}/k_{\mathcal{B}} = 15 $ (see Appendix \ref{app:B}). Figure~\ref{qu2} shows that the corresponding PDF is already very close to a canonical distribution with the same temperature. A fit given by the expression $f(T) = A \exp\left(-{\mathcal{H}_S}/k_{\mathcal{B}}T\right)$, with $A$ as the free parameter and $T$ obtained from Fig.~\ref{fig:dissip}, was applied to the numerical data. One can notice a good agreement between the numerical result and both fits. In this regime, the finite bath is large enough so that it can be almost treated as an usual thermal reservoir. Furthermore, we computed the distribution of angles for $N=10$. The values of $b$ and $c$ obtained form the fit $f(\theta) = b\theta+c $ were essentially the same as in the case of a single QS.

Once we have properly characterized the equilibrium state of the system interacting with the finite bath, we can now study the non-equilibrium fluctuations in this mesoscopic scale.

%%%%%%%%%%%%%%%%%%%%%%%%%%%%%%%%%%%%%%%%%%%%%%%%%%%%%%%%%%%%%%%%%%%%%%%%%%%%%%%%%%%%%%%%%%%%%%%%%%%%%%

\section{ \label{sec:5}Fluctuation Theorem}

The nonequilibrium work statistics of our model is constructed following Ref.~\cite{campisi_finite}. After equilibration, the HO is decoupled from the heat bath and the spring constant $k$ is varied according to a given protocol $k(t)$ between times $t_i$ and $t_f$. Since the HO is isolated, the work performed on the system in each realization of the protocol is $\mathcal{H}_{S}(z_{f},k_{f}) - \mathcal{H}_{S}(z_{i},k_{i})$, where $z_{i,f}$ and $k_{i,f}$ are the phase-space points and spring constant values at $t_{i,f}$. In particular, $z_f = z(t_f,t_i,z_i)$ is the solution of Hamilton's equations of motion with initial condition $z_i$ and the protocol $k(t)$. The work distribution in this process is thus given by the expression
\begin{align}
   \nonumber \rho_F(\mathcal{W}) = \frac{1}{\mathcal{N}_i}\int dz_i\, &\delta(\mathcal{H}_{S}(z_{f},k_{f}) - \mathcal{H}_{S}(z_{i},k_{i}) - \mathcal{W})\\
   \times &\left[1 - \left(\frac{\mathcal{H}_{S}(z_{i},k_{i}) - U_{S,i}}{C_BT_i}\right) \right]_+^{\frac{C_B}{k_\mathcal{B}} -1},
    \label{cuya} 
\end{align}

\noindent where $U_{S,i}$ is the initial value of the internal energy, $T_{i}=T(U_{S,i},k_{i})$ is the initial temperature, defined by Eq.~\eqref{eq:US} and $\mathcal{N}_{i}=\mathcal{N}(U_{S,i},k_{i})$ is given by Eq.~(\eqref{eq:norm}). For simplicity of notation, we label all quantities by a subscript $i$ or $f$, depending on whether the quantity is taken at initial or final values of $k$ and $z$.

Similarly, the work distribution for the reverse protocol reads

\begin{align}
    \nonumber\rho_R(\mathcal{-W}) = \frac{1}{\mathcal{N}_f}\int & dz_f\, \delta(\mathcal{H}_{S}(z_i,k_i) - \mathcal{H}_{S}(z_f,k_f) + \mathcal{W}) \\
   \times &\left[1 - \left(\frac{\mathcal{H}_{S}(z_f,k_f) - U_{S,f}}{C_BT_f}\right)\right]_+^{\frac{C_B}{k_\mathcal{B}} -1},  
    \label{cuya1}
\end{align}

\noindent where now $z_i= z(t_i,t_f,\overline{z}_f)$ is the solution of Hamilton's equations of motion with initial condition $\overline{z}_f = (Q_{f},-P_{f})$ and the time reversal of the protocol $k(t)$. In this case, the work done on the system in each realization of the protocol is $\mathcal{H}_{S,i} - \mathcal{H}_{S,f}$. 

Analogously to Crooks' fluctuation theorem, we can express the ratio between $\rho_F(\mathcal{W})$ and $\rho_R(\mathcal{-W})$  as \cite{campisi_finite} 
 
    \begin{align}
	       \frac{\rho_F(\mathcal{W})}{\rho_R(-\mathcal{W})} = \left(\frac{T_f}{T_i}\right)^{\frac{C_B}{k_\mathcal{B}}-1}e^{\Delta \mathcal{S}/k_\mathcal{B}},
	       \label{yuyuyu}
    \end{align}

\noindent where $\mathcal{S}$ is the thermodynamic entropy in the ensemble \eqref{eq:eq_est} given by \cite{campisi_finite} (see also Appendix~\ref{app:C})  
    \begin{align}
        \mathcal{S} = k_\mathcal{B}\ln{[\mathcal{N}(U_S,k)]}.
        \label{eq:entropia_finit}
    \end{align}
    
\noindent The ratio  \eqref{yuyuyu} is solely written in terms of quantities relative to the equilibrium states of the system. As we will see, the final temperature $T_f$ contains the work dependence on the right side of that equation. In fact, considering the first law of thermodynamics for the whole system, we have that $\Delta U_{B} + \Delta U_{S} = \mathcal{W}$. Thus, we have  
    
    \begin{align}
         T_f = T_i + \frac{\mathcal{W}}{(C_B + k_\mathcal{B})},
    \end{align}
 
\noindent since $U_{B} = C_{B}T$ and $U_S  = k_\mathcal{B}T$ for the HO (see Appendix \ref{app:B}). Additionally, the term in the exponential appearing in Eq.~(\ref{yuyuyu}) is the entropy change $\Delta\mathcal{S}$ defined in \eqref{eq:entropia_finit}. As shown in Appendix \ref{app:C}, the entropy variation can be expressed in terms of the temperatures in the following way. The normalization function for the HO is 
    \begin{align}
            \mathcal{N}(U_S,k) =  \left( \frac{5}{3}\right)^{\frac{1}{2}} \frac{2\pi}{\sqrt{k}} k_\mathcal{B}\,T(U_S).
    \end{align}

\noindent Thus, we obtain the following expression for the variation in entropy between two equilibrium states induced by a finite bath,

 \begin{align}
        \frac{\Delta\mathcal{S}}{k_{\mathcal{B}}} =  \ln{\left[\frac{T_f}{T_i}\left(\frac{k_i}{k_f}\right)^{\frac{1}{2}}\right]}.
        \label{91}
    \end{align}

\noindent Finally, the analytic expression for the fluctuation theorem of our model reads

\begin{align}
      \frac{\rho_F(\mathcal{W})}{\rho_R(-\mathcal{W})} =  \left(\frac{k_i}{k_f}\right)^{\frac{1}{2}}\left[ 1 + \frac{\mathcal{W}}{(1+C_B/k_{\mathcal{B}})U_{S,i}} \right] ^{C_B/k_{\mathcal{B}}}.
      \label{gr}
\end{align}  

In the following, we will compare this analytical result with numerical simulations of the work distributions for the model \eqref{eq:hamilt}. Essentially, we have estimated numerically the ratio $\rho_F(\mathcal{W})/\rho_R(\mathcal{-W})$. In order to do so, we have considered the following procedure in obtaining the numerical data,

\begin{enumerate}
        \item The HO thermalizes with the finite bath at temperature $ T_i $;
        \item The HO is then decoupled from the bath;
        \item We perform the forward protocol on the HO, changing the spring constant $ k $ from $ k_i $ to $ k_f $. In this process, the work $ \mathcal{W} _F $ is performed;  
        \item HO and bath are placed in contact again until the new equilibrium state at temperature $ T_f = T_f(\mathcal{W}_F) $ is reached;
        \item Again, the HO and the bath are decoupled;
        \item The reverse protocol is performed on the system, changing $ k $ from $ k_f $ to $ k_i $. The work done is $ \mathcal{W} _R $.
    \end{enumerate}

\begin{figure}[!ht]
	\centering
	\includegraphics[width=0.5\textwidth]{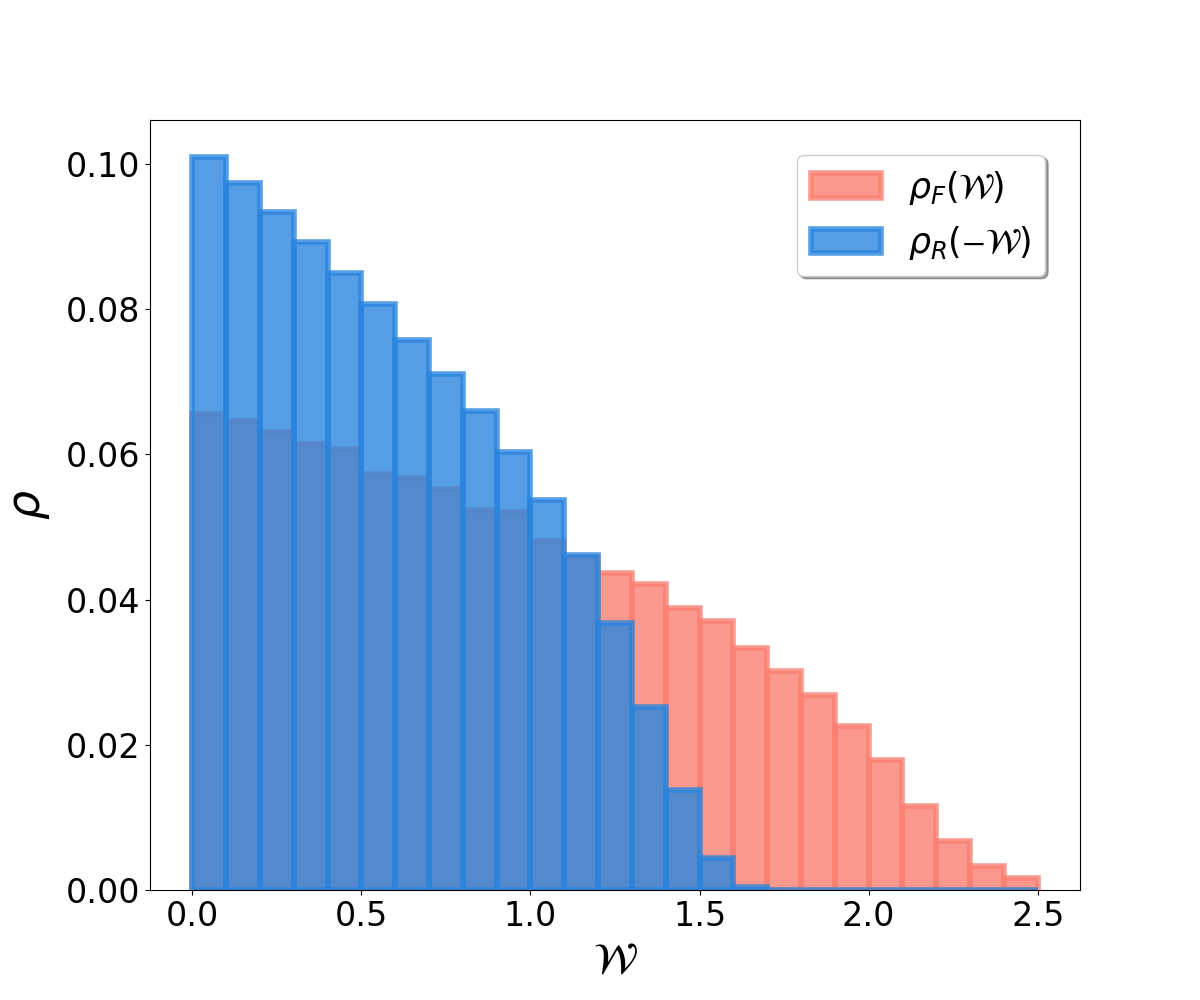}% Here is how to import EPS art
	\caption{Work distributions $\rho_F (\mathcal {W})$ and $\rho_R (\mathcal {-W})$ for the forward and reverse processes, respectively, and for $N=1$. We sampled microcanonically $10^5$ initial conditions $E_S=3.0$, $E_B=10^{-2}$, $a=0.1$, $k = 0.3$, and $\lambda=0.1$. After equilibration, $U_{S,i}=1.23$. The value of $k$ was changed from $0.3$ to $0.9$ within a time interval $\tau_{S}/2$ using a linear protocol.}
	\label{fig:hist_W}
\end{figure}

\begin{figure}[!ht]
	\centering
	\includegraphics[width=0.5\textwidth]{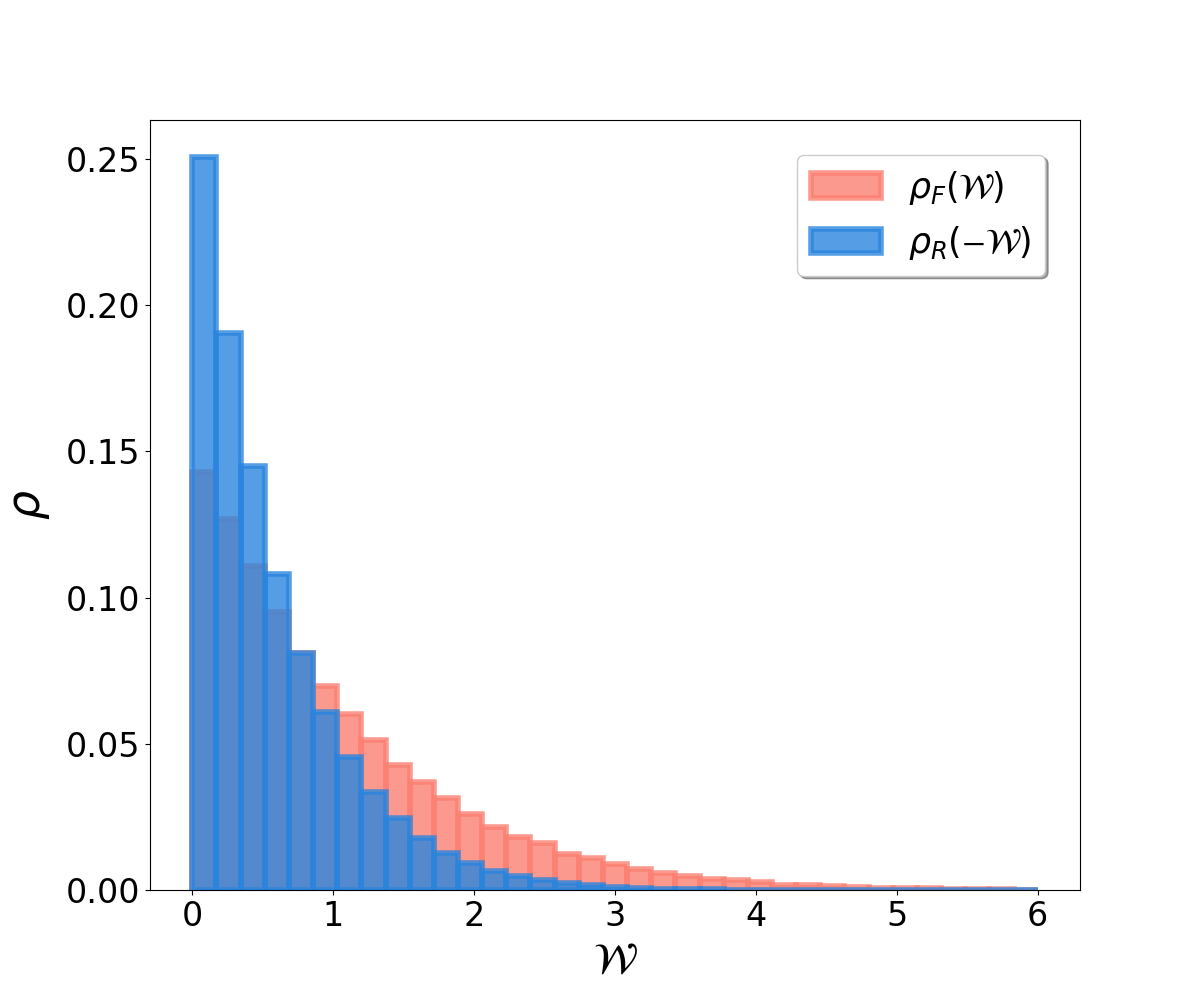}% Here is how to import EPS art
	\caption{Work distributions $\rho_F(\mathcal{W})$ and $ \rho_R(\mathcal{-W})$ for the forward and reverse processes, respectively, and for $N=10$. We sampled microcanonically $10^5$ initial conditions with $E_S=19.7$, $E_B=10^{-5}$, $a=0.1$, $k = 0.3$, and $\lambda=0.1$. After equilibration, $U_{S,i}=1.29$. The value of $k$ was changed from $0.3$ to $0.9$ within a time interval $\tau_{S}/2$ using a linear protocol.}
	\label{fig:hist_W10}
\end{figure}

In order to find $ \rho_F (\mathcal {W}) $, we have integrated numerically the corresponding equations of motion in steps 1 and 3. Two distinct finite baths of model (\ref{eq:hamilt}) were considered, namely, a single QS $(N=1)$ and a collection of ten QS $(N=10)$. After decoupling the HO and the finite bath, we performed a linear protocol with duration $\tau = \tau_S/2 $ taking $k$ from $ k_i = 0.3 $ to $ k_f = 0.9 $. As pointed out before, after the reverse process, the equilibrium state of the system depends on the work done in the forward process. This dependence manifests itself through the equilibrium temperature.  Hence, for different values of work performed on the system in the forward process, it is expected that the system reaches different final equilibrium states after step 4. In other words, the initial equilibrium ensemble splits into several final equilibrium ensembles after the second thermalization step. Thus, steps 5 and 6 were performed for each of these different equilibrium ensembles. After acquiring the corresponding work histograms, we compiled them into a single $ \rho_R (\mathcal {W}) $ which was properly normalized. The work distributions $ \rho_F (\mathcal {W}) $ and $ \rho_R (\mathcal {W}) $ presented in Figs.~\ref{fig:hist_W} and \ref{fig:hist_W10} were obtained from the procedure just described.

%In contrast, caution is required in the calculation of the work distribution for the reverse process $ \rho_R (\mathcal {W}) $, that is, when $ k $ is linearly taken from $ k_f = 0.9 $ to $ k_i = 0.3 $ during same time interval as in forward protocol. It is known that for usual reservoirs that they tend not to be influenced by the coupling with the system of interest. In this regime, the equilibrium state of the system, determined by the canonical distribution, tells us that trajectories with energy less than or equal to the reservoir temperature are more likely than those with higher energy. However, this does not occur for finite baths. Indeed, as pointed out, the equilibrium temperature between the system and the bath is work dependent. That means, for different values of work performed on the system in the forward process, different final equilibrium states can be obtained in step 4. In order to obtain an acceptable numerical estimation, we chose some values of work from the distribution $ \rho_F (\mathcal {W}) $, with which we simulate in the same amount the realizations of the steps 5 and 6. By obtaining the distributions for the reverse protocol, we compile all into a single work distribution $ \rho_R (\mathcal {W}) $, renormalizing it afterward. The work distributions $ \rho_F (\mathcal {W}) $ and $ \rho_R (\mathcal {W}) $ are presented in the figure \ref{fig:hist_W}.

As mentioned before, the work distributions \eqref{cuya} and \eqref{cuya1}, as well as the equilibrium distribution \eqref{eq:eq_est}, have a cutoff energy, which is more evident in Fig.~\ref{qu} and \ref{fig:hist_W}. This leads to spurious results when we take the ratio between the two distributions close to the cutoff value. Hence, we tested various stopping criteria for the calculation of the ratio to avoid this problem. For the histograms shown in Fig.~\ref{fig:hist_W} ($N=1$), the ratio was calculated for the first thirteen intervals of values of work. This is essential for a good comparison with the analytical prediction of the fluctuation theorem.

\begin{figure}[!ht]
	\centering
	\includegraphics[width=0.5\textwidth]{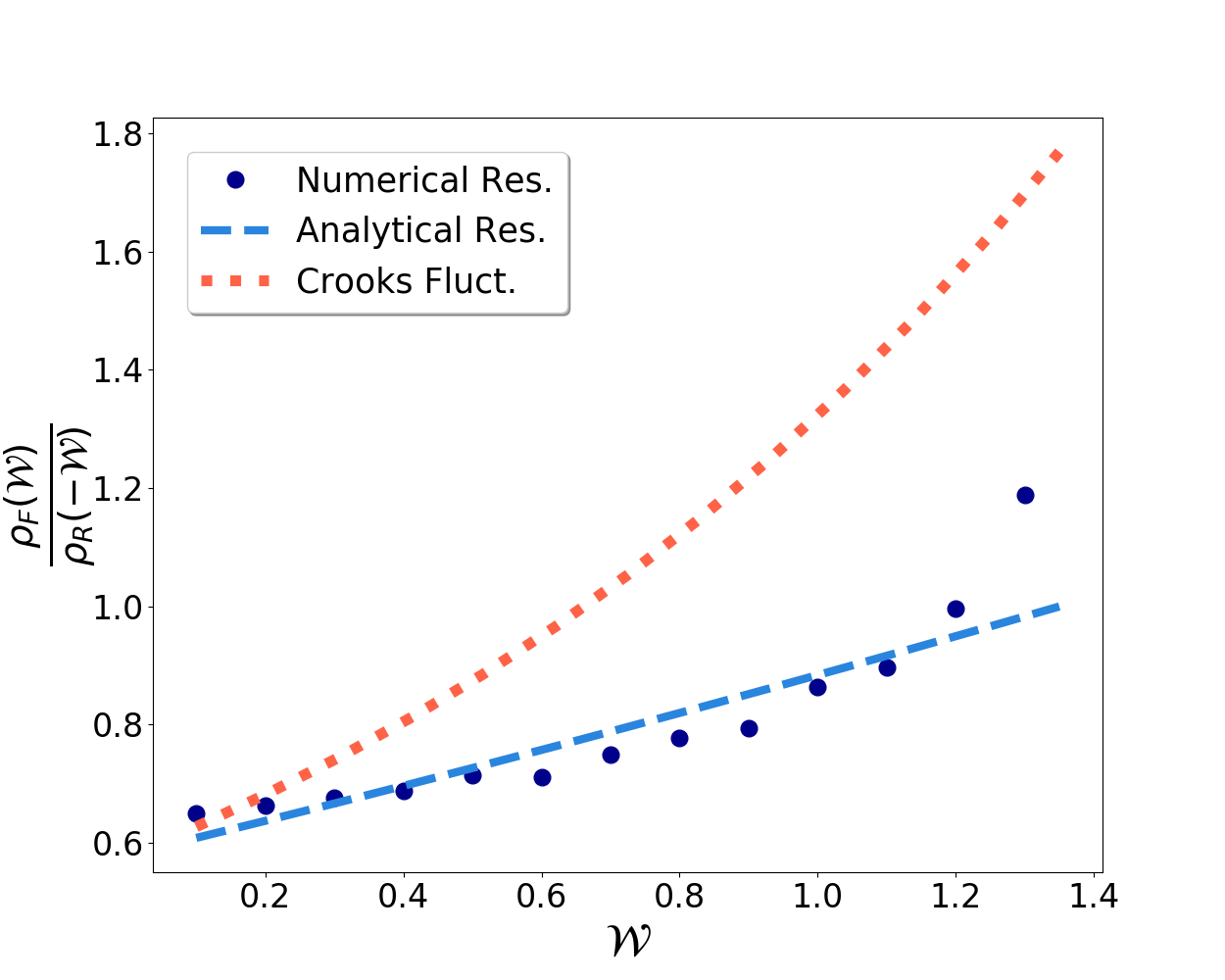}
	\caption{Comparison between analytical (blue dashed line), Eq.~\eqref{gr}, and numerical results (dark blue dots) for the ratio $\rho_{F}(\mathcal{W})/\rho_{R}(\mathcal{-W})$. The numerical results were obtained from the ratio of the histograms shown in Fig. \ref{fig:hist_W} up to the thirteenth interval of work values. The salmon dotted line corresponds to Crooks' fluctuation theorem. We used $U_{S,i}=k_{\mathcal{B}}T_{i}=1.23$, $k_f = 0.9$ and $k_i = 0.3$. The same values were used to calculate the free energy difference $\Delta F$ (see Eq. \eqref{eq:part}). }
	\label{fig:divisao}
\end{figure}

The comparison between numerical and analytical results is presented in Fig.~\ref{fig:divisao}. It is clear that there is a good agreement although the total system has only three degrees of freedom ($N=1$). Since the fluctuation theorem is understood as a detailed information about thermodynamic time asymmetry, our model suggests that this may exist even in small systems. Figure~\ref{fig:divisao} also shows that, for $N=1$, our results are quite distinct from Crooks' fluctuation theorem evaluated for the same values of $k$ and temperature (there is only one temperature in this case and we chose it to be $T_{i}$). Despite of this, a decrease in this distinction is expected as the number of degrees freedom of the bath increases. This is supported by the analysis performed in Ref.~\cite{campisi_finite} about what happens when $C_{B}$ increases. In the limit $C_{B}/k_{\mathcal{B}}\to\infty$, Eq.~(\ref{yuyuyu}) goes to 

\begin{align}
    \frac{\rho_F^{C_B \to \infty}(\mathcal{W})}{\rho_R^{C_B \to \infty}(-\mathcal{W})}  = \frac{ \mathcal{Z}_f(T,k_f)}{ \mathcal{Z}_i(T,k_i)}e^{\beta\mathcal{W}},
\end{align}

\noindent where $\mathcal{Z}(T,k)$ denotes the canonical partition function. For the HO,
\begin{align}
    \frac{ \mathcal{Z}_f(T,k_f)}{ \mathcal{Z}_i(T,k_i)} = e^{-\beta\Delta F}=\sqrt{\frac{k_i}{k_f}}
    \label{eq:part}
\end{align}

\begin{figure}[!ht]
	\centering
	\includegraphics[width=0.5\textwidth]{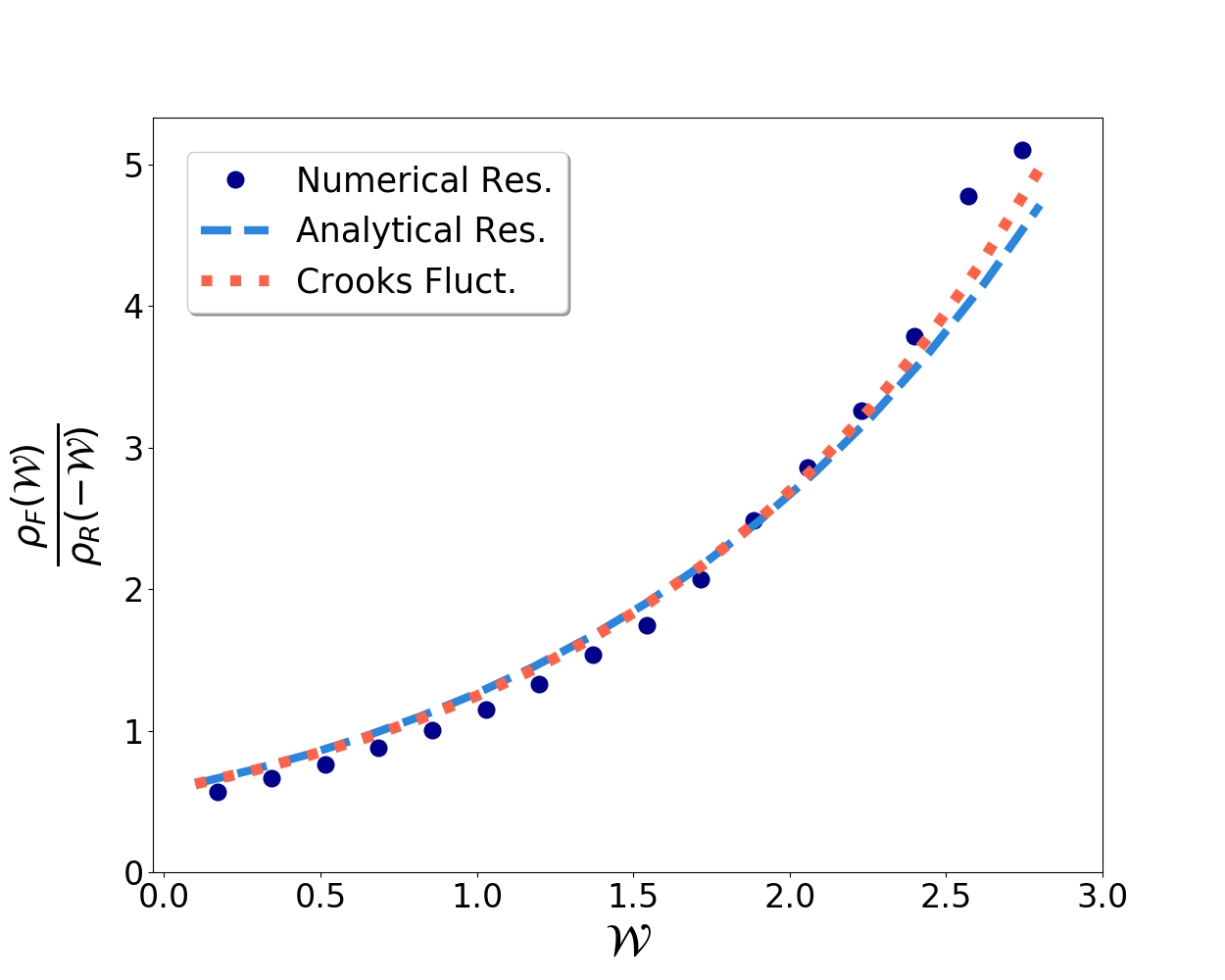}
	\caption{Comparison between analytical (blue dashed line), Eq.~\eqref{gr}, numerical results (dark blue dots) for the ratio $\rho_{F}(\mathcal{W})/\rho_{R}(\mathcal{-W})$. The numerical results were obtained from the ratio of the histograms shown in Fig. \ref{fig:hist_W10} up to the sixteenth interval of work values. The salmon dotted line corresponds to Crooks' fluctuation theorem. We used $U_{S,i}=k_{\mathcal{B}}T_{i}=1.29$, $k_f=0.9$ and $k_i=0.3$. The same values were used in Eq.~\eqref{eq:part}.}
	\label{fig:divisao10}
\end{figure}

Figure~\ref{fig:hist_W10} shows the work distributions obtained numerically for $N=10$ using the same protocol $k(t)$ as in the $N=1$ case. Figure~\ref{fig:divisao10} shows the comparison between numerical results and analytical predictions for the ratio $\rho_{F}(\mathcal{W})/\rho_{R}(-\mathcal{W})$ according to finite bath and Crooks' fluctuation theorems. In this case, the histograms of Fig.~\ref{fig:hist_W10} were divided up to the first sixteen intervals of work values. We observe a very good agreement in this range, even with Crooks' fluctuation theorem.

%%%%%%%%%%%%%%%%%%%%%%%%%%%%%%%%%%%%%%%%%%%%%%%%%%%%%%%%%%%

\section{CONCLUSIONS}

We have shown that the nonequilibrium fluctuations of a certain non-ergodic Hamiltonian system, described by Eqs.~\eqref{eq:hamilt} to \eqref{eq:QS}, follow the fluctuation theorem. Such system is composed of a part of interest (with only one degree of freedom) which is weakly interacting with another part treated as a finite heat bath. A good agreement with the fluctuation theorem was verified even when the bath had a very small number of degrees of freedom, namely, two. In this case, chaotic behavior is already possible if the system lacks two constants of motion in involution \cite{lichtenberg}. As discussed in Secs.~\ref{sec:level2} and \ref{sec:3}, the proposed model allows for a easy control of its dynamical behavior and all results were obtained keeping each sub-system of the heat bath strongly chaotic. Such behavior plays an important role in several aspects of our results, especially the relaxation mechanism. However, there are indeed sets of initial conditions for which the numerical results deviate strongly from the analytical predictions of both equilibrium and nonequilibrium fluctuations even when the chaotic behavior remains. For some of these initial conditions, even the relaxation to an equilibrium state is hardly observed. We have found empirically that the best agreement between analytical and numerical results happens when most part of the total initial energy is given to the system of interest. This can be intuitively understood if we admit that the increment of the energy of each component of the heat bath favors the occupation of non-ergodic parts of the phase space. A more accurate analysis of the dynamical behavior presented by different sets of initial conditions deserves further investigations.

The model we have studied, although non-ergodic and chaotic, allows for an analytical description of the equilibrium distribution induced by the finite heat bath on the system of interest. This was essential for a comparison between numerical results and analytical predictions about the ratio of the nonequilibrium work distributions. We believe that the example we have provided helps to expand the already broad range of applicability of the fluctuation theorem.

\appendix

\section{Averages in the microcanonical ensemble}
\label{app:A}

We show how to calculate microcanonical averages for the QS
Hamiltonian given by 

\begin{align}
    \mathcal{H} = \frac{p_{x}^2 + p_{y}^2}{2} + \frac{x^2 y^2}{2} + a\frac{(x^4 + y^4)}{4}.
    \label{eq:QS1}
\end{align}

\noindent The microcanonical phase-space distribution reads
\begin{align}
    \rho(\mathbf{z},E,a)= \frac{\delta(\mathcal{H}(\mathbf{z},a)-E)}{\int \delta(\mathcal{H}(\mathbf{z},a)-E) \,d\mathbf{z} },
    \label{eq:micro_dis}
\end{align}

\noindent where $\delta(x)$ is the Dirac delta function. For easier manipulation of this distribution, we perform a transformation of variables from $\mathbf{z} = (x,y,p_x,p_y)$ to $(\mathcal{H},\psi,\theta,\phi)$ given by

\begin{align}
    \nonumber x^2 & = \sqrt{\frac{2\mathcal{H}}{\cos 2\theta}}\left(\frac{\cos\theta}{\sqrt{1+a}}+\frac{\sin\theta}{\sqrt{1-a}}\right)\sin\psi,\\
    \label{p1}
    y^2 & = \sqrt{\frac{2\mathcal{H}}{\cos 2\theta}}\left(\frac{\cos\theta}{\sqrt{1+a}}-\frac{\sin\theta}{\sqrt{1-a}}\right)\sin\psi,\\
    \nonumber p_x   & = \sqrt{2\mathcal{H}}\cos\varphi\cos\psi,\\
    \nonumber p_y   & = \sqrt{2\mathcal{H}}\sin\varphi\cos\psi,
\end{align}

\vspace{2mm}

\noindent where $0<\mathcal{H}<\infty, 0<\psi<\frac{\pi}{2}, 0<\theta<\frac{1}{2}\arccos a$, and $0<\varphi<2\pi$. The Jacobian of this transformation $J(\mathcal{H},\psi,\theta,\varphi)$ is defined by the equation,

\begin{align}
    J(\mathcal{H},\psi,\theta,\varphi) = \sqrt{\frac{\mathcal{H}}{2\cos 2\theta( \cos 2\theta - a)}} \cos\psi.
    \label{jacobs}
\end{align}

\noindent  In this case, the microcanonical distribution \eqref{eq:micro_dis} is rewritten as

\begin{align}
    \rho(\mathcal{H},E,a)= \frac{\delta(\mathcal{H}-E)}{\int J(\mathcal{H},\psi,\theta,\varphi) \delta(\mathcal{H}-E) \,d\mathcal{H}\, d\psi\, d\theta\, d\varphi }.
\end{align}

\noindent Thus, the average of any dynamical observable in this ensemble is

\begin{align}
    \nonumber \left< \mathcal{O}\right> &= \int \mathcal{O}(\mathbf{z})\, \rho(\mathbf{z},E,a)\,d\mathbf{z} \\
                              &= \frac{\int \mathcal{O}(E,\psi,\theta,\varphi)\, J(E,\psi,\theta,\varphi) d\psi d\theta d\varphi}{\int J(E,\psi,\theta,\varphi) \, d\psi d\theta d\varphi}.
\end{align}

\noindent In particular, when $\mathcal{O}= K_{QS} =  (p_{x}^2 + p_{y}^2)/2$,  we obtain \begin{align}
    \left< K_{QS} \right> &= E \int_{0}^{\pi/2} \cos{\psi}^3\,d\psi = \frac{2E}{3}.
\end{align}

%%%%%%%%%%%%%%%%%%%%%%%%%%%
\section{Phase-space volume, DOS and microcanonical heat capacity for the QS}
\label{app:B}

The phase-space volume enclosed by a surface of constant energy of the QS Hamiltonian \eqref{eq:QS1} is given by the expression
\begin{align}
\Omega_{QS}(E,a) &= \int \Theta \left(E - \mathcal{H}(\mathbf{z},a) \right) d\mathbf{z},
\label{eq:b1}
\end{align}
where $\Theta(x)$ denotes the Heaviside step function. Using the transformation of variables $\mathbf{z} = (x,y,p_x,p_y) \to (\mathcal{H},\psi,\theta,\varphi)$ (see Eqs.~\eqref{p1}) and considering the Jacobian \eqref{jacobs}, we obtain 
\begin{align}
    \Omega_{QS}(E,a) = \frac{16\pi}{3} \sqrt{\frac{2}{1-a}} F\left( \sin^{-1}\sqrt{\frac{1-a}{1+a}} \left| \frac{1+a}{1-a} \right.\right) E^{3/2},
    \label{eq:volume_esp-fase}
\end{align}
where $F(x_{0}|b)=\int_{0}^{x_{0}}dx (1-b^{2}\sin^{2}{x})^{-1/2}$ is the incomplete elliptic integral of the first kind. Equation~(\ref{eq:volume_esp-fase}) implies that $\Omega_{QS}(E,a) \sim E^{3/2}
$. By differentiating $\Omega_{QS}(E,a)$ with respect to $E$, one obtains the density of states
\begin{align}
    \omega_{QS}(E,a) = 8\pi \sqrt{\frac{2}{1-a}} F\left( \sin^{-1}\sqrt{\frac{1-a}{1+a}} \left| \frac{1+a}{1-a} \right.\right)E^{1/2}.
    \label{eq:DOS1}
\end{align}

%%%%%%%%%%%%%%%%%%%%%%%%%%%

%\section{Microcanonical heat capacity for many QS's}
%\label{app:C}

%In this appendix, we show how to calculate the microcanonical heat capacity of a set of quartic systems (QS's) each of them described by Hamiltonian $\mathcal{H}_{QS}$  
%\begin{align}
%mathcal{H}_{QS} = \frac{p_{x}^2 + p_{y}^2}{2} + \frac{x^2 y^2}{2} + a\frac{(x^4 + y^4)}{4}
%\end{align}

We show now how to calculate the microcanonical heat capacity of a set of QS each of them described by Hamiltonian \eqref{eq:QS1}. For a single QS, the microcanonical heat capacity is given by the 
    \begin{align}
        C_{QS} = \left( \frac{\partial T_{QS}}{\partial E}\right)^{-1},
        \label{eq:heat_cap}
    \end{align}
    
\noindent where $T_{QS} = \left(\partial \mathcal{S}_{QS}/\partial E\right)^{-1} =\Omega_{QS}/ \omega_{QS}$ and $\mathcal{S}_{QS} = k_\mathcal{B}\ln \Omega_{QS} $. With the help of expressions \eqref{eq:volume_esp-fase} and \eqref{eq:DOS1}, we obtain

    \begin{align}
        C_{QS} = C_{B}^{(1)} = \frac{3}{2}k_\mathcal{B}.
        \label{eq:heat_cap1}
    \end{align}

For $N$ quartic systems, the phase-space volume $\Omega_{B}$ is obtained through the Eq.~\eqref{eq:bath_conv} recalling that $\mathcal{H}_{B}=\sum_{i=1}^{N}\mathcal{H}_{B_{i}}$ where each $\mathcal{H}_{B_{i}}$ is given by Eq.~\eqref{eq:QS1}. Thus, from expressions \eqref{eq:volume_esp-fase} and \eqref{eq:DOS1}, we obtain for $N=3$

\begin{align}
    \Omega_{B}(E) &\propto \int_{0}^{E} d\mathcal{H}_{B_{1}} \int_{0}^{E-\mathcal{H}_{B_{1}}} d\mathcal{H}_{B_{2}}\nonumber\\
    &\times\mathcal{H}_{B_{1}}^{1/2} \mathcal{H}_{B_{2}}^{1/2} \left(E-\mathcal{H}_{B_{1}}-\mathcal{H}_{B_{2}} \right)^{3/2} \propto E^{9/2}.
\end{align}

\noindent It is possible to show then by induction that, for arbitrary $N$, $\Omega_{B}(E)\propto E^{N\times 3/2}$. Together with Eq.~\eqref{cuenver}, this implies that $C_{B}/k_{\mathcal{B}} = N\times 3/2 = N C_{B}^{(1)}/ k_{\mathcal{B}}$.

The phase-space volume $\Omega_{S}$ is defined analogously to Eq.~\eqref{eq:b1} with the Hamiltonian \eqref{eq:QS1} replaced by HO Hamiltonian $\mathcal{H}_{S}$ given by Eq.~\eqref{eq:HO}. It is easy to show then that
\begin{align}
    \Omega_{S}(E) = \int \Theta(E-\mathcal{H}_{S}(z,k)) dz\propto E.
\end{align}
\noindent Hence, Eq.~\eqref{cuenver} leads to $C_{S}/k_{\mathcal{B}}=1$.

%%%%%%%%%%%%%%%%%%%%%%%%%%%
\section{HO Internal energy an entropy for finite bath}
\label{app:C}

In this appendix, we calculate the internal energy and the entropy for a system of interest, when it is in contact with a finite bath. This system is a HO described by the Hamiltonian $\mathcal{H}_S$, Eq.~\eqref{eq:HO}, and its equilibrium state with a finite bath is characterized by distribution \eqref{eq:eq_est}.

The internal energy $ U_S $ of the system is given by
    \begin{align}
           U_S \doteq \langle \mathcal{H}_S(z,k) \rangle, 
            \label{eq:media1}
    \end{align}

\noindent where $z = (Q,P)$. Equation~\eqref{eq:media1} can be rewritten as
\begin{align}
   \frac{1}{\mathcal{N}}\int  dz\, ( \mathcal{H}_S  - U_S )\left[ 1- \left(\frac{\mathcal{H}_S(z,k) - U_S}{C_BT}\right) \right]_{+}^{\frac{C_B}{k_\mathcal{B}} -1} = 0.
    \end{align}

\noindent The following change of variables is considered,
\begin{align}
    Q = \sqrt{2\mathcal{H}_S/k} \,\cos{\phi},\;\;
    P = \sqrt{2\mathcal{H}_S} \,\sin{\phi},
    \label{eq:param_HO}
\end{align}

\noindent  where $0<\mathcal{H}_S<E_{tot} = U_S + C_BT$, $\,0<\phi<2\pi\,$ and the Jacobian is $J(\mathcal{H}_S,\phi) = 1/\sqrt{k}$. Therefore, we obtain
    \begin{align}
            \int_{0}^{U_S + C_BT} &d\mathcal{H}_S \, (\mathcal{H}_S- U_S )   \left[ 1- \left(\frac{\mathcal{H}_S - U_S}{C_BT}\right) \right]^{\frac{C_B}{k_\mathcal{B}} -1} = 0.
            \label{conta}
    \end{align}

\noindent We can integrate Eq.~\eqref{conta} by parts noticing that the following equality holds,
\begin{align}
    -k_\mathcal{B}T\frac{\partial}{\partial \mathcal{H}_S} \bigg[ 1- &\left(\frac{\mathcal{H}_S - U_S}{C_{B}T}\right) \bigg]^{\frac{C_B}{k_\mathcal{B}}} \nonumber \\
    &=\left[ 1- \left(\frac{\mathcal{H}_S - U_S}{C_{B}T}\right) \right]^{\frac{C_B}{k_\mathcal{B}} -1}.
\end{align}

\noindent This leads to the expression,
    \begin{align}
            \left(\frac{C_B}{k_\mathcal{B} + C_B}\right)\left[ 1 + \frac{U_S}{C_BT} \right]^{\frac{C_B}{k_\mathcal{B}}}\left(  U_S  -k_\mathcal{B}T   \right) = 0 .
    \end{align}
\noindent which must hold for any positive values of $U_{S}$, $T$ and $C_{B}$. Thus, for the HO, the following relation applies,
    \begin{align}
        U_S \doteq \langle \mathcal{H}_S(z,k) \rangle = k_\mathcal{B}T.
        \label{eq:linear_dep}
    \end{align}
    
Let us now calculate the entropy for the HO. In the ensemble \eqref{eq:eq_est}, the entropy is given by \cite{campisi_finite}
    \begin{align}
        \mathcal{S} = k_\mathcal{B}\ln{[\mathcal{N}(U_S,k)]},
        \label{eq:entropia_finit1}
    \end{align}

\noindent where $\mathcal{N}(U_S,k)$ is normalization factor (see Eq.~\eqref{eq:norm}). In the equilibrium state with induced by a finite bath composed of a single QS, Eq.~\eqref{eq:QS1}, $\mathcal{N}$ reads
    \begin{align}
            \nonumber \mathcal{N}(U_S,k) &= \frac{2\pi}{\sqrt{k}}\int_{0}^{E_{tot}} \, d\mathcal{H}_S \left[ 1- \left(\frac{\mathcal{H}_S - U_S}{C_BT}\right) \right]^{\frac{C_B}{k_\mathcal{B}}-1}  \\& = \frac{2\pi}{\sqrt{k}} k_\mathcal{B}T \left(1 + \frac{2}{3}\frac{U_S}{k_\mathcal{B}T}\right)^{\frac{1}{2}},
            \label{eq:norm1}
    \end{align}
    \vspace{3mm}
    
\noindent where Eq.~\eqref{eq:heat_cap1}, the change of variables \eqref{eq:param_HO}, and $E_{tot}= U_S + C_B T$ were considered. Using Eq.~\eqref{eq:linear_dep}, we obtain
        \begin{align}
            \mathcal{N}(U_S,k) =  \frac{2\pi}{\sqrt{k}} k_\mathcal{B}T(U_S) \left( \frac{5}{3}\right)^{\frac{1}{2}}.
            \label{eq:norm2}
    \end{align}

\noindent Using Eq.~\eqref{eq:norm2} in Eq.~\eqref{eq:entropia_finit1}, we obtain
        \begin{align}
            \mathcal{S} =  k_\mathcal{B}\ln{\left[\frac{2\pi}{\sqrt{k}} T \left( \frac{5}{3}\right)^{\frac{1}{2}}\right]}.
    \end{align}
    
Hence, the entropy variation between two equilibrium states of the system is
    \begin{align}
       \noindent \Delta\mathcal{S} = \mathcal{S}(&T_f(U_f),k_f) - \mathcal{S}(T_i(U_i),k_i) \\ &= k_\mathcal{B}\ln{\left[\frac{T_f}{T_i}\left(\frac{k_i}{k_f}\right)^{\frac{1}{2}}\right]}.
        \label{eq:vari_entr}
    \end{align}

%%%%%%%%%%%%%%%%%%%%%%%%%%%%%%%%%%%%%%%%%%%%%%%%%%%%%%%%%%%%%%%%%%%%%%%%%%%%%%%%%%%%%%%%%%%%%%%%%%%%%%

%\vspace*{-0.2cm}
\begin{acknowledgments}
\vspace*{-0.2cm}
The authors gratefully acknowledge the support from the Brazilian agency CNPq under Grants 134590/2016-3 and  142556/2018-1. 
%We gratefully acknowledge support from the Brazilian agencies CNPq, Fapesp and specially to Capes for financial support through of the project PROEX - 0487. The calculations were performed at CCJDR-IFGW-UNICAMP.  

\end{acknowledgments}

%%%%%%%%%%%%%%%%%%%%%%%%%%%%%%%%%%%%%%%%%%%%%%%%%%%%%%%%%%%%%%%%%%%%%%%%%%%%%%%%%%%%%%
%\nocite{*}
\bibliographystyle{apsrev4-1}
%\nocite{apsrev41Control}
%\bibliography{bibliografia}% Produces the bibliography via BibTeX.

%

\end{document}